# Crystal structure and basic properties of dirhenate quantum materials

Danrui Ni [a], Xianghan Xu [c], Stephen Zhang [a], N.P. Ong [b], Sanfeng Wu [b] and R. J. Cava* [a]

[a]Department of Chemistry, Princeton University, Princeton, NJ 08544, United States

[b]Department of Physics, Princeton University, Princeton, NJ 08544, United States

[c]School of Physics and Astronomy, University of Minnesota, Twin Cities, MN 55455

* Corresponding Author: Robert J. Cava (E-mail: rcava@princeton.edu)

## Abstract

The anhydrous divalent 3*d*-metal dirhenate quantum materials, ***M***$(ReO_4)_2$,were synthesized using solid-state methods for ***M*** = Mn through Zn. Previously unreported $Mg(ReO_4)_2$ is also described. Their layered crystal structures, which feature an in-plane triangular lattice of $\boldsymbol{M}^{2+}$, were refined using single crystal X-ray diffraction, and their optical absorption and several other physical properties were characterized. Their magnetism and heat capacity reveal long-range magnetic order at low temperatures in many of the ***M***$(ReO_4)_2$ phases. Notably, many of these ordered states are sensitive to applied magnetic fields and can be readily suppressed by relatively small fields, suggesting competing magnetic interactions in a low-dimensional framework, which appear worthy of further study.

## Introduction

Layered magnets have drawn considerable research attention since the discovery of superconductivity in a layered cuprate [1], and as potential realizations of multiple quantum state models with interesting spin correlations and spin-orbit entanglement [2,3]. Magnets with layered triangular motifs have been the subject of much interest in current condensed matter physics due to this motif's ability to host strongly-correlated magnetic moments and geometrically frustrated spins. In the search for atypical phenomena and exotic behavior, such as spin glasses, quantum spin liquids, and unconventional superconductors [4–6], this lattice type has attracted considerable interest. The quantum materials in this system are worthy of study, especially when they are based on transition metals that have strong magnetic exchange interactions, as the spin state in such systems serves as a fertile ground for exploring the rich physics of strongly correlated electron systems [7,8].

The anhydrous divalent metal rhenates $\boldsymbol{M}(ReO_4)_2$ with their layered crystal structures are one such system of research interest. Their unit cells are constructed from $\boldsymbol{M}^{2+}$O6 octahedra and $Re^{7+}$O4 tetrahedra , with an $\boldsymbol{M}^{2+}$-based triangular lattice in-plane, positioning them as prospective candidates for hosting unconventional phenomena and spin correlations [9]. These materials, especially with divalent transition metals as $\boldsymbol{M}^{2+}$ (Mn through Zn), have been discovered previously, prepared by the direct solid state synthesis and solution synthesis methods or dehydration of the hydrates [9,10]. However their structural information remains scarce, as high-precision refinements remain lacking, and even the assignment of their space group may be controversial for some formulas (for example, Torardi et al. used *P*-3*m*1 symmetry for $\boldsymbol{M}(ReO_4)_2$ with ***M*** = Mn/Co/Ni [9], while Butz et al. reported *P*-3 for these compounds based on their powder diffraction patterns [10]). Moreover, systematic investigations of their magnetism are still absent.

In this report, the anhydrous divalent metal rhenates $\boldsymbol{M}(ReO_4)_2$ (***M*** = Mg/Mn/Fe/Co/Ni/Cu/Zn) for many 3*d* elements are directly prepared via a solid-state synthesis method. High-quality refinements of their crystal structures are carried out with single crystal X-ray diffraction (SCXRD). Their structural properties, optical absorption, and magnetic behaviors are characterized and analyzed, as well as the variation with different ***M***. Their properties highlight the rich magnetic behaviors arising from layered triangular magnetic lattices and underscore their potential as candidate quantum materials for future research.

## Experimental

Anhydrous divalent metal rhenates $\boldsymbol{M}(ReO_4)_2$ ($\boldsymbol{M}$ = Mg/Mn/Fe/Co/Ni/Cu/Zn) were synthesized into polycrystalline powders using solid-state methods. Metal oxides, metal carbonates, or metal dichlorides were used as starting materials as the ***M*** metal source, mixed with rhenium metal (Alfa Aesar, 99.99%) and annealed at 400 – 500 °C in air. After annealing for 2 days, samples were quickly transferred to the glove box chamber and quench cooled from the annealing temperature under vacuum, in order to prevent absorbing moisture during cooling. Due to the hygroscopicity of the materials, all the rhenate compounds were handled in an air-free atmosphere, and their powder X-ray diffraction (PXRD) patterns were collected with a Rigaku Miniflex II diffractometer located inside a nitrogen-filled glove box, using Cu Kα radiation ($\lambda$ = 1.5406 Å). Le Bail refinements were conducted using Topas software. The results showed that the carbonates and chlorides generally had better reactivity and yielded purer products than the oxides did, as the metal oxides ($\boldsymbol{M}O_x$) appeared to be the most common impurities in the post-reaction samples. Adding more Re and reannealing at the same temperature improved the purity of the products when needed.

Attempts to grow single crystals of anhydrous $\boldsymbol{M}(ReO_4)_2$ were conducted using different methods and with different transport agents. $TeCl_4$ was one of the most helpful agents when using chemical vapor transport in vacuum-sealed quartz tubes with temperature gradients (Small amounts of metal chlorides were observed, but can be distinguished from the target products.). $Fe(ReO_4)_2$ crystals of mm size were successfully grown using this method with 450 °C at the hot zone. For the other transition metals, heating the samples at 500 – 550 °C for longer time increased the crystallinity and particle sizes, resulting in small single crystals that could be isolated. For $Cu(ReO_4)_2$, annealing the sample at 500 °C resulted in melting, but small single crystals could be separated from the melted chunk. Single Crystal X-ray diffraction (SCXRD) was performed at 295 K under a nitrogen gas flow, using a Bruker D8 Quest Eco diffractometer equipped with a Photon III CPAD detector and monochromatic Mo Kα radiation ($\lambda$ = 0.71073 Å). The crystal structure refinements were performed by using the SHELXTL Software Package [11,12].

Scanning electron microscopy (SEM) images together with energy-dispersive X-ray spectroscopy (EDX) characterizations were collected using a Quanta 200 FEG environmental scanning electron microscope equipped with an Oxford EDX system for elemental analysis. UV–vis–NIR diffuse reflectance spectra were collected on polycrystalline sample powders (mixed and diluted with $BaSO_4$ at a 1:10 weight fraction), using an Agilent Cary 5000 spectrometer, with an Agilent Internal DRA-2500 diffuse reflectance accessory. An O-ring sealed sample holder was employed to prevent sample hydration during the measurements, and $BaSO_4$ was used as the blank for baseline correction. The diffuse reflectance data were converted to pseudo absorption by using the Kubelka–Munk function and a Tauc plot [13,14] was used to estimate the possible band gap values.

Magnetization and heat capacity measurements were conducted, with efforts made to minimize contact of the hygroscopic materials with the air. The experiments were carried out on a Quantum Design PPMS (Dynacool), equipped with a vibrating sample magnetometer (VSM). Magnetic susceptibility was defined as M/H, and the temperature-dependent magnetization (M) was measured in an applied magnetic field (H) of 1000 Oe on powder samples, and on single crystals for $Fe(ReO_4)_2$. Powder samples were cold pressed into pellets sufficiently dense for heat capacity measurements, except for $Fe(ReO_4)_2$ whose heat capacity measurement was conducted on a piece of grown single crystals. For these and all other cases, transfers were performed very rapidly to prevent hydrate formation.

## Results and discussion

Polycrystalline powders of anhydrous $\boldsymbol{M}(ReO_4)_2$ were successfully prepared by a direct solid-state method. After several attempts to grow single crystals of anhydrous $\boldsymbol{M}(ReO_4)_2$ using different methods and with different transport agents, high-quality SCXRD refinement results were obtained for all seven compounds at ambient temperature. (**Table 1**, and **Table S1-S3**). Their unit cell is refined to have a layered structure (**Figure 1**), and Le Bail fittings were conducted on their PXRD patterns to confirm the consistency of the bulk materials with the crystals. (**Figure 2** and **Figure S1-S5**). Images from optical microscope and SEM revealed the hexagonal layered morphology of the particles (**Figure 3**), while EDX of SEM confirmed the 1:2:8 elemental ratio for all the compounds, consistent with the expected formulas (see the table in Figure 3).

The refined two-dimensional structure, in *P*-3*m*1 symmetry, can be viewed through Figure 1A. Each block, stacked along the *c*-axis, can be considered as a single plane of ***M***O6 octahedra sandwiched by two layers of ReO4 tetrahedra. In the *ab*-plane, the ***M***O6 octahedra form a large triangular lattice with a distance of ~ 5.7 to 5.9 Å (the parameter *a* of the unit cells), which are corner-connected but only by ReO4 tetrahedra. Each oxygen atom in an ***M***O6 octahedron is bonded to a separate Re atom; there is no direct ***M***-O-***M*** bonding. This loose in-plane triangular lattice of $M^{2+}$ ions is reflected in the low temperatures of their transitions to magnetically ordered states. The unshared apical O atom in each ReO4 tetrahedron "interpenetrates" the adjacent layers.

$Cu(ReO_4)_2$, different from the other compounds in this report, was refined to have a monoclinic *C*2/*m* unit cell by SCXRD at 300 K (Figure 1B), which can be considered as distorted from the *P*-3*m*1 structure, with twisting and tilting of the metal-oxide polyhedra. Although for this compound the Jahn-Teller distortion of the $Cu^{2+}$O6 octahedra must be the driving force for this effect, this structure type was also reported for the low-temperature phase of $Zr(MoO_4)_2$ [15]. The relationship between the two structures can be more clearly viewed by the orange-dashed frames in Figure 1C, and thus we can compare the dimensions of $\sqrt{3}a$ & *a* in the *P*-3*m*1 unit cell, versus the *a* and *b* parameters for $Cu(ReO_4)_2$ in its *C*2/*m* cell.  The CuO6 octahedra are obviously more elongated compared to the sister compounds in the *P*-3*m*1 structure (Table S2 and S3). This introduces a strong Jahn-Teller effect, and is consistent with the tendency of forming square-planar coordination for $Cu^{2+}$. It is also noted that there are several extra peaks with very low intensity found in the PXRD patterns (labeled by the orange stars in Figure 2B) which cannot be matched in the PXRD database. However, by enlarging the current *C*2/*m* unit cell (for example, to double *c*), these very low intensity peaks can be fitted. This suggests that these peaks may come from a superstructure or sublattice with a new periodicity, which is present in a small fraction of the sample, and thus is not detected by SCXRD.

Lattice parameters *a* and *c* of ***M***$(ReO_4)_2$ (for ***M*** = Cu, due to the different unit cell symmetry, "*a*" is taken as the averaged edge length of the in-plane $Cu^{2+}$ triangle, and *c* is the *c* of its *C*2/*m* unit cell), are plotted in Figure 2C, following a left-to-right sequence of ***M*** in the periodic table. The variation of the lattice parameter *a* is consistent with the variation of the ionic radius of $M^{2+}$. The lattice parameter *c* does not vary substantially, except for Cu, whose *c* is obviously smaller than others and drops from the trend. For the *P*-3*m*1 structure compounds (ignoring Mg, which is not in the series), the parameter *c* generally increases with increasing atomic number. The volume of the ***M***O6 octahedron and the average ***M***-O bond length ($d_{M\text{-}O}$) are plotted together with the ionic radius of $M^{2+}$ in the right panel of Figure 2C, showing consistent variation. $Ni^{2+}$ is revealed to be at the bottom of the group, exhibiting the most compact ***M***O6 octahedron and shortest ***M***-O distance, as well as the shortest ***M***-***M*** distance ($d_{M\text{-}M}$ = the lattice parameter *a* of the *P*-3*m*1 unit cell) of the in-plane triangular lattice.

Different from the *P*-3 structure reported for the refinement of powder X-ray diffraction data [10], ***M***$(ReO_4)_2$ (***M***=Mg/Mn/Fe/Co/Ni/Zn) was refined with a P-3m1 unit cell by SCXRD, as the oxygen atom on the 6i site (O1 in the crystallographic tables) is found to align with the metal atoms, preserving the mirror symmetry perpendicular to the *ab*-plane (**Figure S6**). This symmetry is consistent with the observations of a somewhat earlier paper [9]. A comparison was then conducted by fitting the PXRD patterns with different structure models (with *P*-3*m*1 and *P*-3 symmetries), and it was found that the *P*-3*m*1 unit cell consistently gave a slightly better refinement covariance, while the *P*-3 model always refined the O1 atom to have a larger $U_{iso}$ value. This further supports the *P*-3*m*1 symmetry for the 300 K structure determined by SCXRD for the series.

Additionally, the alkaline earth metal rhenate, $Mg(ReO_4)_2$ has the P-3m1 structure, which may reveal the effect of metal cation size of $M^{2+}$ in ***M***$(ReO_4)_2$ structure type. Other divalent main-group metals with larger atomic sizes adopt different structures for their rhenates. For example, Ca and Sr, are reported to have a *P*21/*n* symmetry unit cell [16], while $Ba(ReO_4)_2$ reportedly crystallizes in *P*31*m* symmetry, isostructural with $Pb(ReO_4)_2$ [17,18], with the face-sharing polyhedra stacking into chains, which are connected by ReO4 tetrahedra by shared oxygen atoms (**Figure S7**). The ionic radius of the $\boldsymbol{M}^{2+}$ ion appears to be limited for this group, as we find that for the seven divalent metal cations in this report, the radii range from 0.69 to 0.83 Å [19,20], while a wide electron count is allowed. For the divalent metal ions with radius larger than 1.0 Å (including $Ca^{2+}$, $Sr^{2+}$, $Ba^{2+}$, and $Pb^{2+}$), a higher coordination number (8~9) is found.

To better understand the properties of the $\boldsymbol{M}(ReO_4)_2$ compounds, diffuse reflectance (DR) spectra were collected on the polycrystalline samples using an integrating sphere. The DR spectra are plotted in **Figure S8**, revealing that most of the absorption features of the $\boldsymbol{M}(ReO_4)_2$ samples were observed in the UV-visible range. The reflectance data were transferred into pseudo absorbance using Kubelka-Munk function (**Figure 4**). Band-gap-like behavior can be observed for the $\boldsymbol{M}(ReO_4)_2$ samples with ***M*** = Ni, Cu, Zn, and Mg. An estimation of the possible band gap values was therefore conducted using a Tauc plot (**Figure S9**). All four samples, if the observed absorptions are caused by a band gap transition, will be insulators with a band gap larger than 3 eV. This would explain their white or light colors.

There are also other absorption peaks observed in Figure 4 for some $\boldsymbol{M}(ReO_4)_2$ samples, which may be from *d-d* transitions for the $\boldsymbol{M}^{2+}$ions, potentially explaining the colors of the $\boldsymbol{M}(ReO_4)_2$ compounds (listed in the table of Figure 3). For example, the peak at around 430-440 nm for $Ni(ReO_4)_2$ yields yellow as a complementary color, which can result in a light-yellow colored crystal, and the light-beige colored powders. These are consistent with our observations, and indicate that $Ni(ReO_4)_2$ is a ferromagnetic insulator (see below) with a relatively large band gap.

Magnetic susceptibility (M/H) was measured under a magnetic field of 1000 Oe, both zero-field-cooled (ZFC) and field-cooled (FC), from 1.8 to 300 K for each sample. When ***M*** = Mg and Zn, diamagnetic behavior for most of the temperature range (**Figure S10**) is observed, with a small paramagnetic upturn at very low temperature. The diamagnetism is consistent with expectations based on their electron configurations, while the weak low temperature upturn may come from trace amounts of magnetic impurities in the samples. $Cu(ReO_4)_2$ shows paramagnetic behavior, with no transition or deviation between ZFC & FC observed in the measured temperature range (**Figure 5**). For ***M*** = Mn/Fe/Co/Ni, a transition below 10 K can be observed in the MT curve with a ZFC-FC deviation (**Figure 6-9**). Anisotropic measurements were conducted on single crystals of $Fe(ReO_4)_2$, with magnetic field direction perpendicular to the layers (out-of-plane) and parallel to the layers (in-plane) respectively (**Figure 9**). In the low temperature range, the $Fe(ReO_4)_2$ crystal shows much stronger magnetic response with the applied field in-plane than out-of-plane. This is consistent with the magnetism measured on the polycrystalline $Fe(ReO_4)_2$ powders (**Figure S11**).

Curie-Weiss fitting was then conducted with Equation (1) in the high temperature range of the MT curves (mostly from 150-300 K), as shown in **Figure S12** for each magnetic compound.

$$\chi - \chi_0 = \frac{C}{T-\theta} \qquad (1)$$

Their fitting results, as well as the measured transition temperatures in magnetism ($T_A$, obtained respectively from the peaks in $d\chi/dT$ curves.) are summarized in **Table 2**, with the effective moments calculated as $\mu_{\mathrm{eff}} = \sqrt{8C}$. Based on the Curie-Weiss θ, the materials tend to be antiferromagnetic (AFM) for ***M*** = Mn/Fe/Co (negative θ), and ferromagnetic for Cu and especially Ni. The Mn and Co samples with negative θ may preserve a small degree of magnetic frustration, as the absolute values of the fitted θ are generally higher than the $T_A$ observed; For $Fe(ReO_4)_2$, Curie-Weiss fitting gives different θ values for in-plane and out-of-plane measurements, and the experimentally measured $T_A$ falls between the two θs.

$\mu_{\mathrm{eff}}$ is plotted versus different transition metal electron configurations in this structure type with the comparison to the calculated spin-only moment $\mu_s$ ($\mu_s = g\sqrt{S(S+1)}$, where $S$ is the total spin of the paramagnetic center and $g$ ≈ 2.00) in **Figure 10**. It is revealed that Mn, Ni, and Cu seem to be close to the spin-only model, while Co and Fe have larger deviation with a $\mu_{\mathrm{eff}}$ higher than $\mu_s$. This suggests that the Co and Fe rhenate have stronger spin-orbit coupling present than the other members of this family.

Magnetization was also measured on varying the applied magnetic field from -9 T to 9 T at 2 K and 250 K. Most of the samples show an essentially linear variation at 250 K and an S-shaped-variation at 2 K. For $Cu(ReO_4)_2$ the 2 K M-H curve shows an S-shape with a saturation at around 1 $\mu_B$ at high magnetic field, which is consistent with the expectation for $d^9$ configuration $Cu^{2+}$. Other compounds also have M-H curves at 2 K that saturate at the expected values, such as $Co(ReO_4)_2$ saturating at around 2 $\mu_B$. $Mn(ReO_4)_2$, on the other hand, shows more linear behavior at 2 K, which suggests a high saturation field limit for this compound, consistent with the high-spin $d^5$ configuration (with five unpaired electrons) for $Mn^{2+}$. For ***M*** = Co and Fe, a very weak hysteresis can be observed in their 2 K loop, while $Ni(ReO_4)_2$ shows a much more obvious opening of its M-H curve (Figure 8), whose shape and behavior are consistent with the expectations for ferromagnetism. This makes $Ni(ReO_4)_2$ a candidate for future research interest, as it potentially is an insulator presenting predominantly ferromagnetic interactions between the magnetic moments on a triangular lattice.

Heat capacity data were then collected on the magnetic ***M***$(ReO_4)_2$ samples. A plate single crystal was used for ***M*** = Fe, with applied magnetic field penetrating the crystal plane; for ***M*** = Mn/Co/Ni/Cu, polycrystalline powder-pressed pellets or as-made dense pieces were used for the measurements. As described **in Figure 11-14**, with ***M*** = Mn/Co/Ni/Fe, a λ-shaped anomaly below 10 K was observed in their heat capacity behavior under zero magnetic field, which corresponds to the transitions seen in the magnetic susceptibility measurements (Table 2) and confirms the long-range nature of the magnetic orders. This is the only transition observed in the range of 2-100 K for these rhenate samples, however considering their structure type and related polymorphic analogs (like $Zr(MoO_4)_2$ [10,15]), a potential structural phase transition cannot be ruled out below room temperature. For these 4 samples, the observed peak in heat capacity was broadened and shifted to higher temperature by the applied magnetic field. Given the layered structure with triangular lattice of ***M***$(ReO_4)_2$, there may be some spin fluctuations and certain degrees of frustration in the system, however the longer $d_{M\text{-}M}$ (5.7 to 5.9 Å) of the loose triangular lattice may also weaken the spin interactions. The anomaly can be completely suppressed when applying a field at 6-9 T for ***M*** = Co/Ni/Fe, while for $Mn(ReO_4)_2$, the broadening and the suppression of the λ-peak by the field is limited. That suggests the presence of a more robust ordered state in $Mn(ReO_4)_2$. For ***M*** = Cu, no obvious anomaly was observed in its heat capacity between 2 and 100 K (Figure S13). This is consistent with its paramagnetic susceptibility behavior down to 1.8 K. However, there is a subtle indication for an upturn below 2 K in the heat capacity curve, which suggests a transition at an even lower temperature. Therefore, very-low-temperature heat capacity measurements were conducted in the PPMS from 0.7 to 10 K using a $^3He$ refrigerator insert (**Figure 15**). A λ-peak was observed at around 1.3 K, which shows similar variation with its sister dirhenates under an applied magnetic field (broadened, suppressed, and shifted to higher temperature). Thus, long-range magnetic order at 1.3 K for $Cu(ReO_4)_2$ is possible, making this compound a promising platform for exploring quantum magnetism. Fitting of phonon contribution was conducted using a modified Debye equation (Eq (2) [21]), on the heat capacity data from 15-100 K, a higher temperature range to avoid the potential magnetic contributions (**Figure S14**). The fitted atomic ratio is close to the expected 3:8 heavy-to-light element ratio for the suggested formula ($\boldsymbol{M}_1Re_2O_8$).

$$C_{phonon} = 9R \sum_{n=1}^{2} C_n \left(\frac{T}{\Theta_{Dn}}\right)^3 \int_0^{\Theta_{Dn}/T} \frac{x^4 e^x}{(e^x-1)^2} dx \qquad (2)$$

After subtracting the fitted phonon contribution, the remaining heat capacity is considered as the magnetic contribution ($C_{mag} = C_{total} - C_{phonon}$), and the $C_{mag}/T$ is plotted versus temperature in the B Panels of Figure 11-14. ΔS is calculated from the integration of the peak in $C_{mag}/T$ and compared to the values of $R\ln(2S+1)$ for local moments. For most of the samples, the values are relatively close to $R\ln(2S+1)$ (suggesting that most entropy change from their magnetic ordering can be found and the materials can be fully magnetic ordered) [22], except for Co, whose ΔS is much lower than the Heisenberg model with high spin state ($R\ln(2S+1)$ with $S = 3/2$), but is closer to the $S = 1/2$ spin with $R\ln(2)$. This suggests that the $Co^{2+}$ in the triangular lattice of $Co(ReO_4)_2$ has an effective $S = 1/2$ Ising ground state, ~~in~~ other words that it may have a spin-orbit entangled pseudospin environment, where strong spin-orbit coupling mixes the spin and orbital states isolating the localized the $J_{eff}$ = 1/2 Kramers doublet as the ground state [23]. Consistently, as shown in Figure 10, the effective moment for $Co^{2+}$ in $Co(ReO_4)_2$ is higher than the spin-only $\mu_S$, which identifies this compound as a promising candidate and motivates future studies of quantum magnetism.

## Conclusions

The anhydrous dirhenate ***M***$(ReO_4)_2$ quantum materials with ***M*** = Mg/Mn/Fe/Co/Ni/Cu/Zn were successfully prepared by solid-state methods. Their crystal structures were refined by SCXRD to have *P*-3*m*1 symmetry for ***M*** = Mg/Mn/Fe/Co/Ni/Zn, and *C*2/*m* symmetry for Cu, which is distorted from the *P*-3*m*1 structure. The consistency of the bulk samples with these cells was confirmed by PXRD. The materials have a layered crystal structure with an in-plane triangular lattice of $\boldsymbol{M}^{2+}$, which leads to rich magnetic behaviors. A long-range magnetic order is revealed for ***M*** = Mn/Fe/Co/Ni, with transition temperatures below 10 K, by magnetism and heat capacity measurements; similarly, a transition at 1.3 K was detected by $^3$He heat capacity measurement for ***M*** = Cu. Curie-Weiss fitting results and magnetization measurements suggest the dominance of AFM correlations for Mn/Fe/Co, and predominantly ferromagnetic correlations for $Ni^{2+}$and $Cu^{2+}$. The current property characterization suggests that the triangular layered ***M***$(ReO_4)_2$ materials appear to be a promising quantum materials family for the future study of rich and competing magnetic interactions in triangular lattices.

## Supplemental Material

The Supporting Information in [24] contains single-crystal crystallographic data, additional PXRD patterns with Le Bail fits, diffuse reflectance spectra and Tauc plots, additional magnetic susceptibility data, Curie−Weiss analysis, and heat capacity data with Debye fitting.

## Appendix

CCDC deposition numbers 2558905-2558911 contain the crystallographic data, which can be obtained free of charge.

## Acknowledgements

This research was primarily funded by the U.S. Department of Energy, Office of Science, National Quantum Information Science Research Centers, Co-design Center for Quantum Advantage (C2QA) under grant DE-SC0012704. S.W. acknowledges support from Office of Naval Research through award N000142612220 and the Gordon and Betty Moore Foundation through Grant GBMF11946. X.X. acknowledges the support by the U.S. Department of Energy through the University of Minnesota Center for Quantum Materials under Grant No. DE-SC-0016371. The authors acknowledge the use of the Imaging and Analysis Center (IAC) operated by the Princeton Materials Institute at Princeton University, which is supported in part by the Princeton Center for Complex Materials (PCCM), a National Science Foundation (NSF) Materials Research Science and Engineering Center (MRSEC; DMR-2011750).

## Conflicts of interest

The authors declare that they have no known competing financial interests or personal relationships that could have appeared to influence the work reported in this paper.

## Data Availability

The data that support the findings of this study are available from the corresponding author upon reasonable request.

**Table 1.** Atomic coordinates and equivalent isotropic displacement parameters for ***M***$(ReO_4)_2$ (***M*** = Mg, Mn, Fe, Co, Ni, Zn, Cu) at 295 K. ($U_{eq}$ is defined as one-third of the trace of the orthogonalized $U_{ij}$ tensor (Å$^2$)). The standard deviations are indicated by the values in parentheses. The occupancy of each atom is 1 on each Wyck. site.

P-3m1 $Mg(ReO_4)_2$:

| Atom | Wyck. | *x* | *y* | *Z* | $U_{eq}$ |
|---|---|---|---|---|---|
| Mg1 | 1*a* | 0.00000 | 0.00000 | 0.00000 | 0.0153(9) |
| Re1 | 2*d* | 0.33333 | 0.66667 | 0.28181(5) | 0.01272(19) |
| O1 | 6*i* | 0.1701(3) | 0.3403(7) | 0.1908(6) | 0.0296(10) |
| O2 | 2*d* | 0.33333 | 0.66667 | 0.5576(10) | 0.0249(15) |

P-3m1 $Mn(ReO_4)_2$:

| Atom | Wyck. | *x* | *y* | *Z* | $U_{eq}$ |
|---|---|---|---|---|---|
| Mn1 | 1*a* | 0.00000 | 0.00000 | 0.00000 | 0.0208(4) |
| Re1 | 2*d* | 0.33333 | 0.66667 | 0.28950(5) | 0.01682(17) |
| O1 | 6*i* | 0.1737(3) | 0.8263(3) | 0.1978(6) | 0.0395(10) |
| O2 | 2*d* | 0.33333 | 0.66667 | 0.5679(12) | 0.0390(17) |

P-3m1 $Fe(ReO_4)_2$:

| Atom | Wyck. | *x* | *y* | *Z* | $U_{eq}$ |
|---|---|---|---|---|---|
| Fe1 | 1*a* | 0.00000 | 0.00000 | 0.00000 | 0.0170(5) |
| Re1 | 2*d* | 0.66667 | 0.33333 | 0.71461(6) | 0.0149(2) |
| O1 | 6*i* | 0.3441(8) | 0.1720(4) | 0.8073(7) | 0.0346(12) |
| O2 | 2*d* | 0.66667 | 0.33333 | 0.4381(12) | 0.034(2) |

P-3m1 $Co(ReO_4)_2$:

| Atom | Wyck. | *x* | *y* | *Z* | $U_{eq}$ |
|---|---|---|---|---|---|
| Co1 | 1*a* | 0.00000 | 0.00000 | 0.00000 | 0.0148(3) |
| Re1 | 2*d* | 0.33333 | 0.66667 | 0.28304(4) | 0.01310(15) |
| O1 | 6*i* | 0.6593(6) | 0.8297(3) | 0.1911(5) | 0.0320(8) |
| O2 | 2*d* | 0.33333 | 0.66667 | 0.5613(9) | 0.0303(13) |

P-3m1 $Ni(ReO_4)_2$:

| Atom | Wyck. | *x* | *y* | *Z* | $U_{eq}$ |
|---|---|---|---|---|---|
| Ni1 | 1*a* | 0.00000 | 0.00000 | 0.00000 | 0.0121(5) |

| Re1 | 2*d* | 0.66667 | 0.33333 | 0.28057(9) | 0.0105(2) |
|---|---|---|---|---|---|
| O1 | 6*i* | 0.8316(6) | 0.6631(11) | 0.1880(10) | 0.0284(13) |
| O2 | 2*d* | 0.66667 | 0.33333 | 0.5576(15) | 0.023(2) |

P-3m1 $Zn(ReO_4)_2$:

| Atom | Wyck. | *x* | *y* | *Z* | $U_{eq}$ |
|---|---|---|---|---|---|
| Zn1 | 1*a* | 0.00000 | 0.00000 | 0.00000 | 0.0169(4) |
| Re1 | 2*d* | 0.33333 | 0.66667 | 0.28250(6) | 0.0145(2) |
| O1 | 6*i* | 0.6590(8) | 0.8295(4) | 0.1907(8) | 0.0368(12) |
| O2 | 2*d* | 0.33333 | 0.66667 | 0.5571(13) | 0.0333(19) |

C2/m $Cu(ReO_4)_2$:

| Atom | Wyck. | *x* | *y* | *Z* | $U_{eq}$ |
|---|---|---|---|---|---|
| Cu1 | 2*a* | 0.00000 | 0.00000 | 0.00000 | 0.0255(3) |
| Re1 | 4*i* | 0.15240(3) | 0.50000 | 0.28150(5) | 0.01737(16) |
| O1 | 8*j* | 0.0628(4) | 0.2464(8) | 0.2069(8) | 0.0377(11) |
| O2 | 4*i* | 0.2966(6) | 0.50000 | 0.1546(12) | 0.0357(15) |
| O3 | 4*i* | 0.1817(9) | 0.50000 | 0.5619(11) | 0.053(2) |

**Table 2.** The magnetic anomaly temperature ($T_A$) from magnetic susceptibility (MT) and heat capacity (HC) measurements, as well as the effective moment ($\mu_{\mathrm{eff}} = \sqrt{8C}$) and Curie-Weiss temperature (θ) obtained from Curie-Weiss fitting, for magnetic ***M***$(ReO_4)_2$ samples.

| ***M*** of ***M***$(ReO_4)_2$ | $T_{A\,(MT)}$ (K) * | $T_{A\,(HC)}$ (K) | $\mu_{\mathrm{eff}}$ | θ (K) |
|---|---|---|---|---|
| **Mn** | 3.2 | 3.0 | 6.1 | -11.6 |
| **Fe** | 8.9 | 8.3 | 6.9 | -6.9 (out-of-plane)<br>-14.8 (in-plane) |
| **Co** | 5.2 | 5.0 | 5.4 | -26.8 |
| **Ni** | 12.7 | 11.9 | 2.7 | +20.8 |
| **Cu** | - | 1.3 | 2.0 | +2.3 |

* Obtained according to the peaks in dχ/dT curves

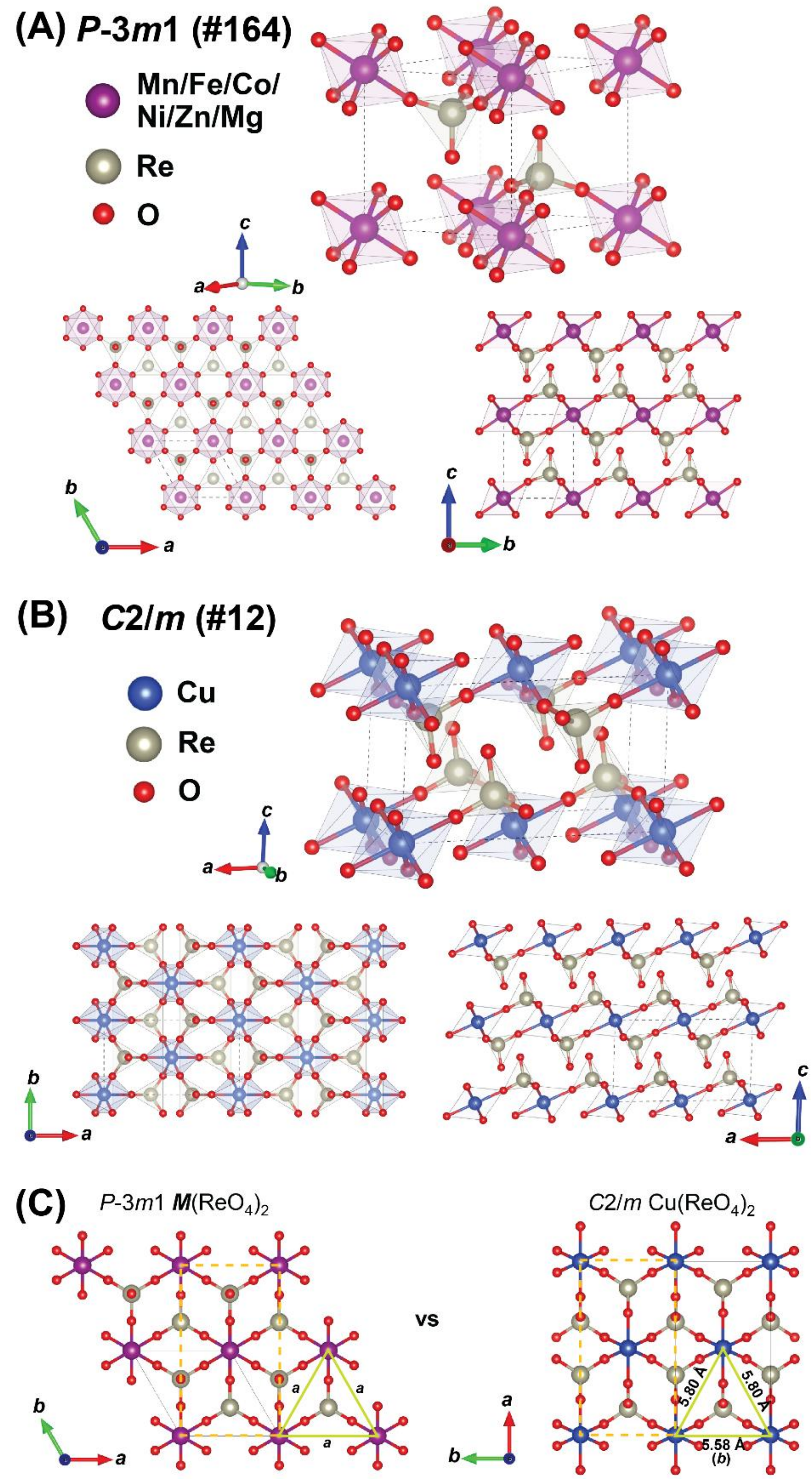

**Figure 1.** The crystal structures of $\boldsymbol{M}(ReO_4)_2$, with views through different axes respectively, in (A) *P*-3*m*1 for ***M*** = Mg/Mn/Fe/Co/Ni/Zn and (B) *C*2/*m* for ***M*** = Cu. A comparison view is shown in (C) with the orange dashed line showing the relationship between the two unit cells.

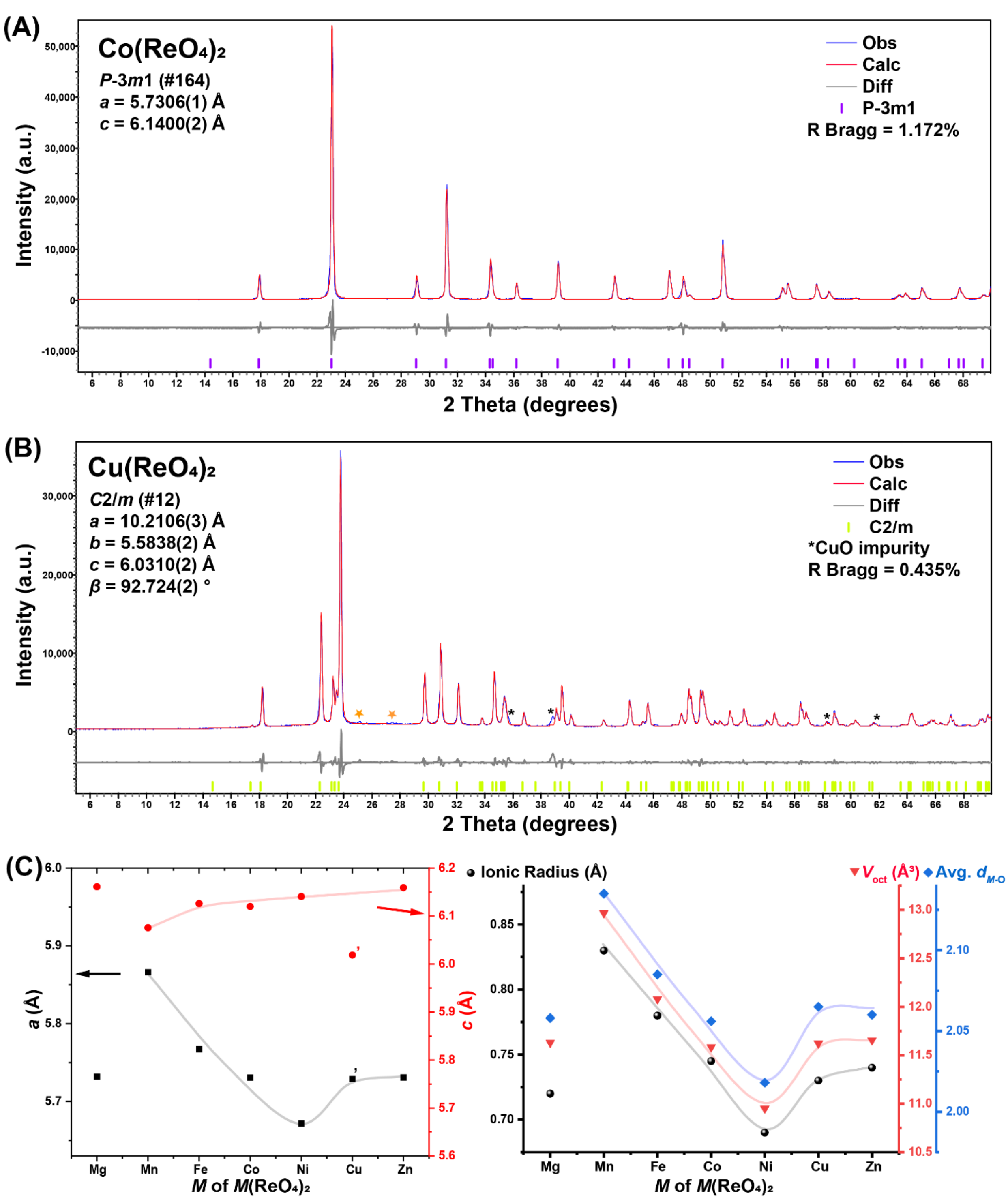

**Figure 2.** Representative PXRD patterns with Le Bail fits for $\boldsymbol{M}(ReO_4)_2$, with (A) ***M*** = Co and (B) ***M*** = Cu. Lattice parameter variations with different metals ***M*** are plotted in the left panel of (C) (* labels the *C*2/*m* structured $Cu(ReO_4)_2$, whose *a'* is the average edge length of the in-plane $Cu^{2+}$ triangle, and *c'* is the *c* of its *C*2/*m* unit cell). The $\boldsymbol{M}^{2+}$ ionic radius, ***M***O6 octahedron volume, and average ***M***-O bond length are plotted on the right.

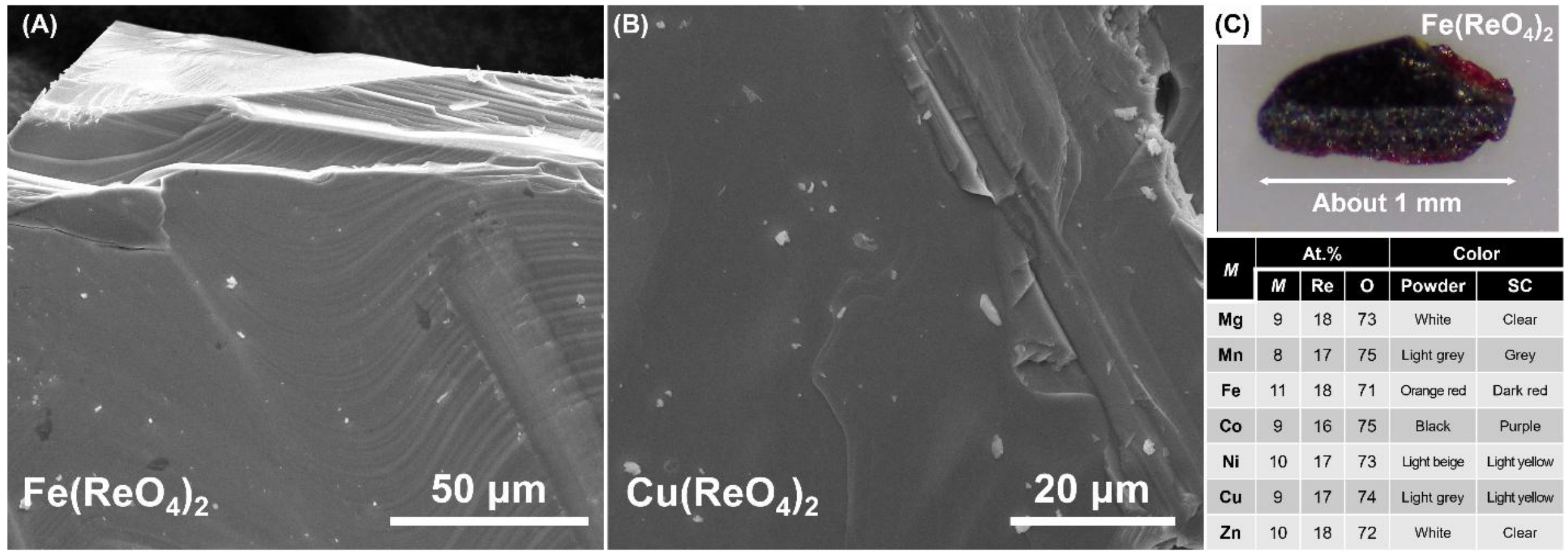

| *M* | At.% | | | Color | |
|---|---|---|---|---|---|
| | *M* | Re | O | Powder | SC |
| Mg | 9 | 18 | 73 | White | Clear |
| Mn | 8 | 17 | 75 | Light grey | Grey |
| Fe | 11 | 18 | 71 | Orange red | Dark red |
| Co | 9 | 16 | 75 | Black | Purple |
| Ni | 10 | 17 | 73 | Light beige | Light yellow |
| Cu | 9 | 17 | 74 | Light grey | Light yellow |
| Zn | 10 | 18 | 72 | White | Clear |

**Figure 3.** Representative SEM images of (A) $Fe(ReO_4)_2$ and (B) $Cu(ReO_4)_2$ samples showing their layered structures; The EDX results of all seven rhenate samples are summarized in the right-bottom table, as well as the colors of the samples. (C) Digital photo taken by an optical microscope on a small piece of a vapor-transport-grown $Fe(ReO_4)_2$ crystal. The inserted table exhibits the EDX results and color information of ***M***$(ReO_4)_2$ samples.

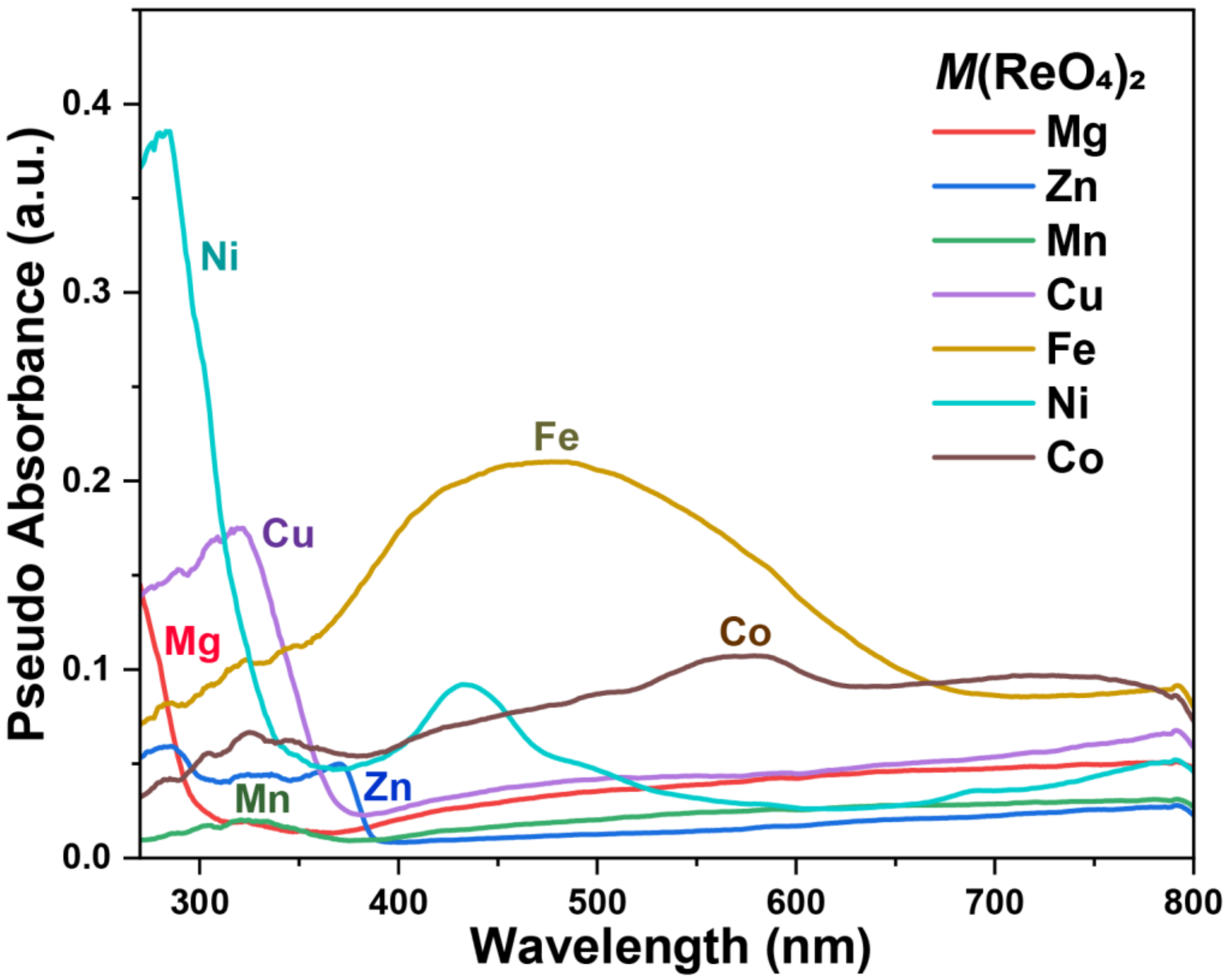

**Figure 4.** Pseudoabsorbance spectra of ***M***$(ReO_4)_2$ powder samples, measured by diffuse reflectance spectroscopy.

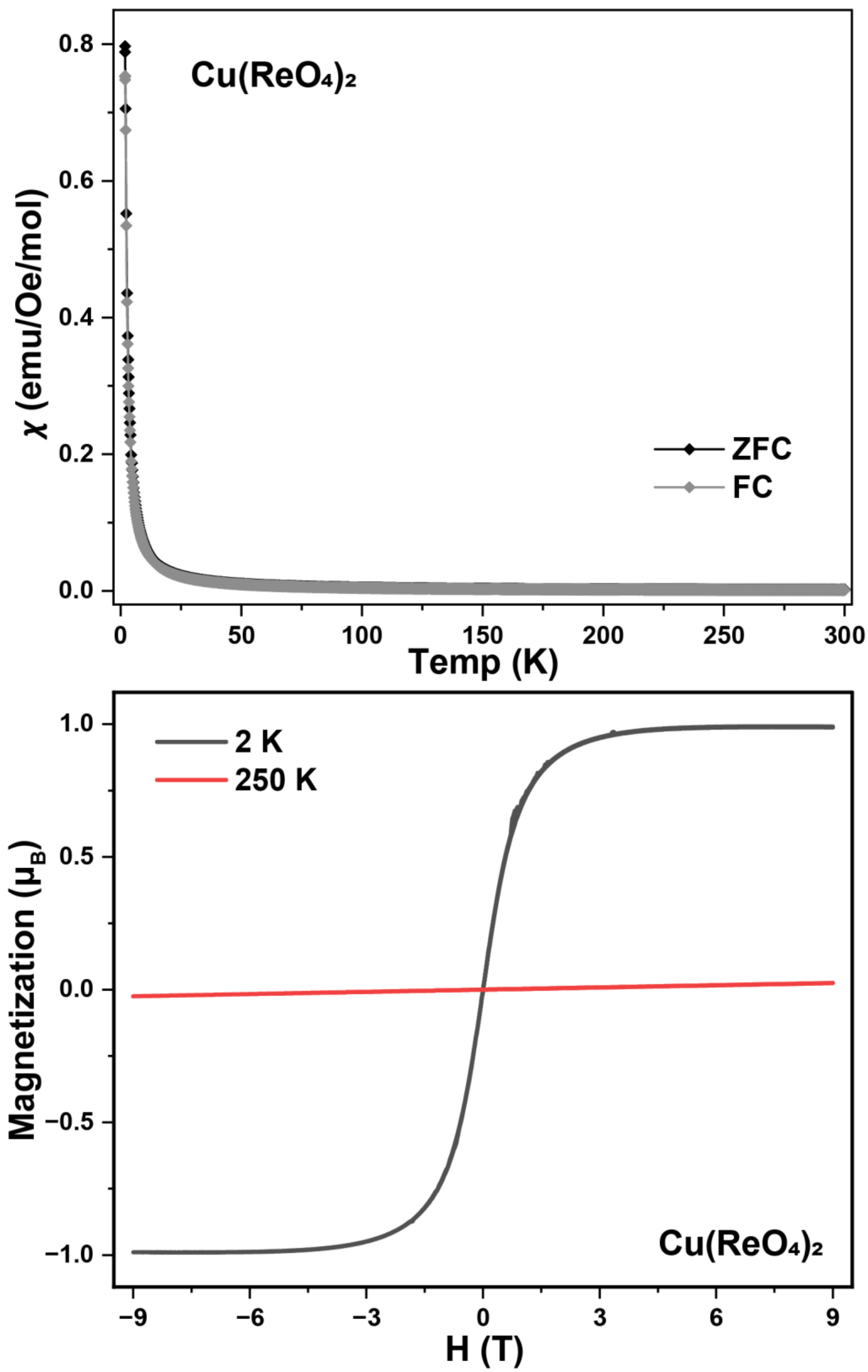

**Figure 5.** The magnetic susceptibility measured from 1.8 to 300 K under 1000 Oe (ZFC & FC) for $Cu(ReO_4)_2$ polycrystalline powders (top), together with the field-dependent magnetization data collected at 2 K and 250 K (bottom).

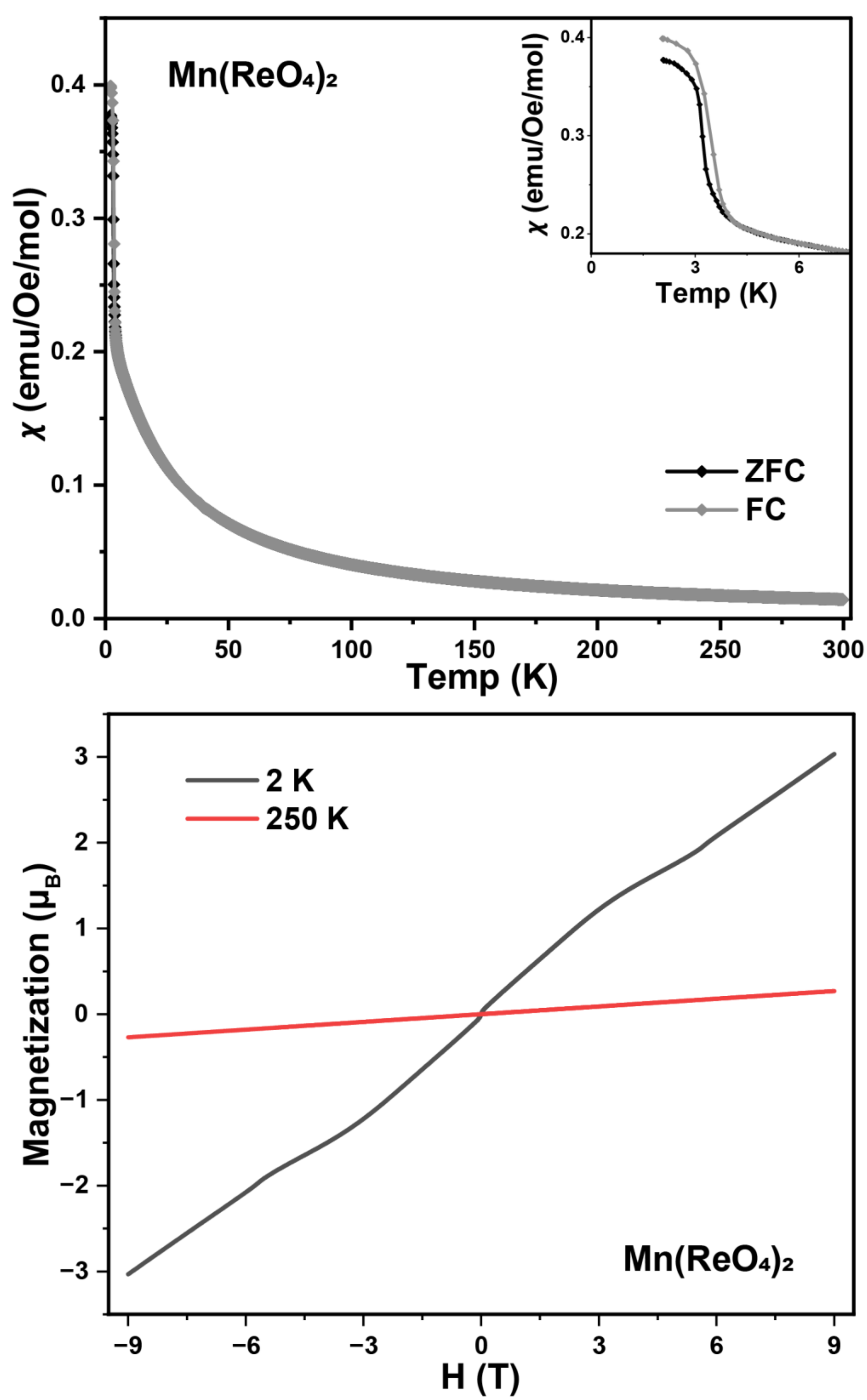

**Figure 6.** The magnetic susceptibility measured from 1.8 to 300 K under 1000 Oe (ZFC & FC) for $Mn(ReO_4)_2$ polycrystalline powders (top), together with the field-dependent magnetization data collected at 2 K and 250 K (bottom).

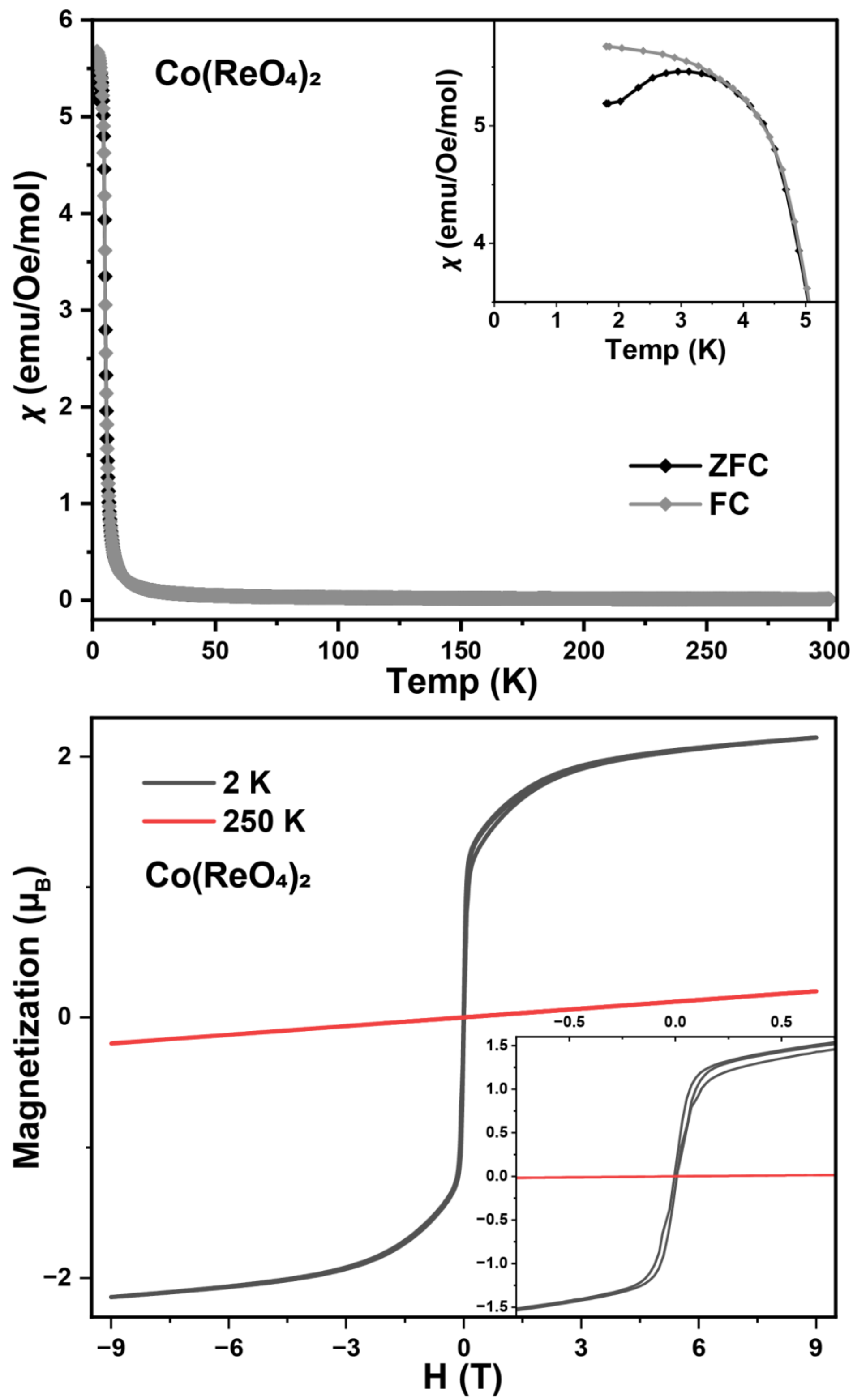

**Figure 7.** The magnetic susceptibility measured from 1.8 to 300 K under 1000 Oe (ZFC & FC) for $Co(ReO_4)_2$ polycrystalline powders (top), together with the field-dependent magnetization data collected at 2 K and 250 K (bottom).

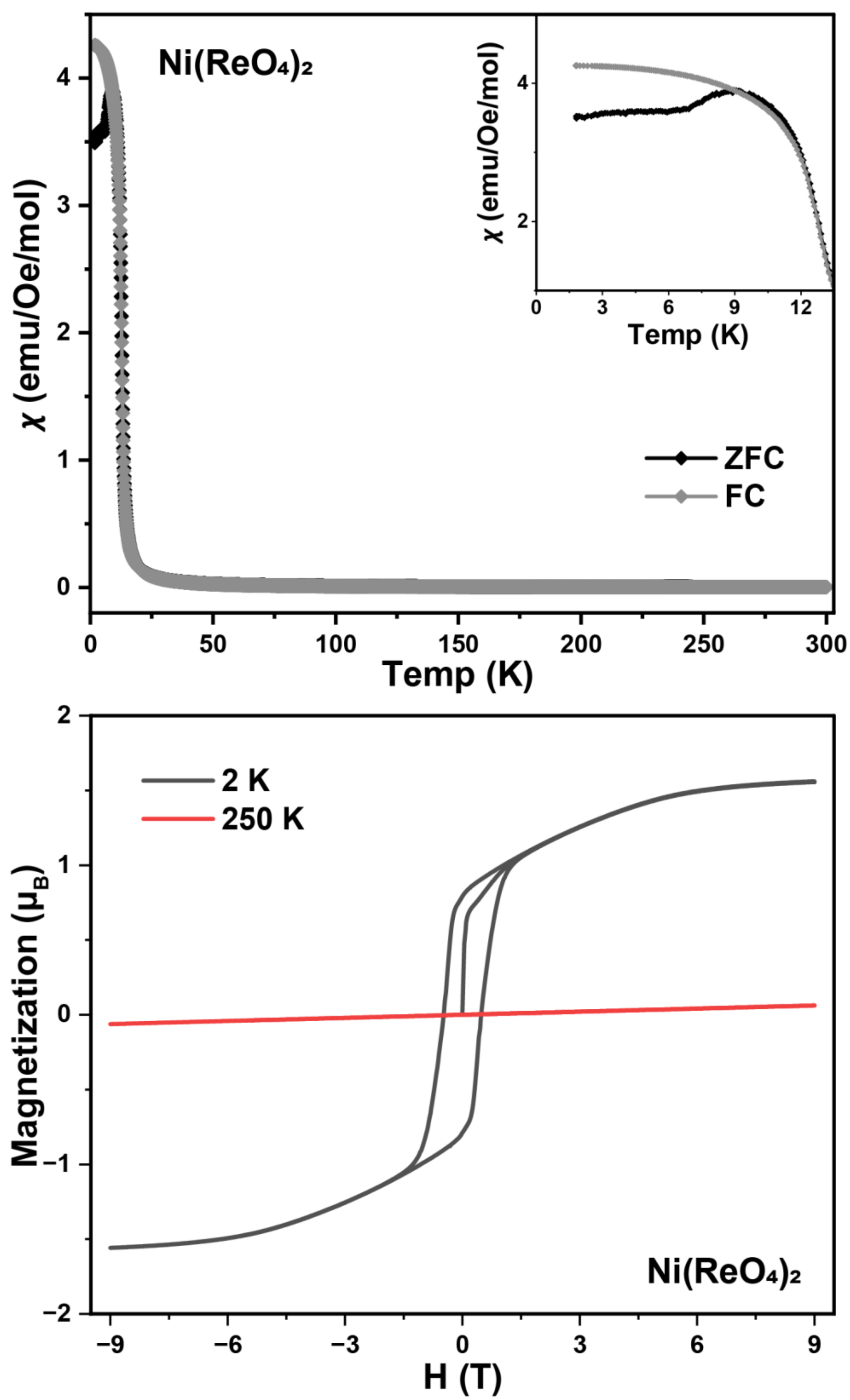

**Figure 8.** The magnetic susceptibility measured from 1.8 to 300 K under 1000 Oe (ZFC & FC) for $Ni(ReO_4)_2$ polycrystalline powders (top), together with the field-dependent magnetization data collected at 2 K and 250 K (bottom).

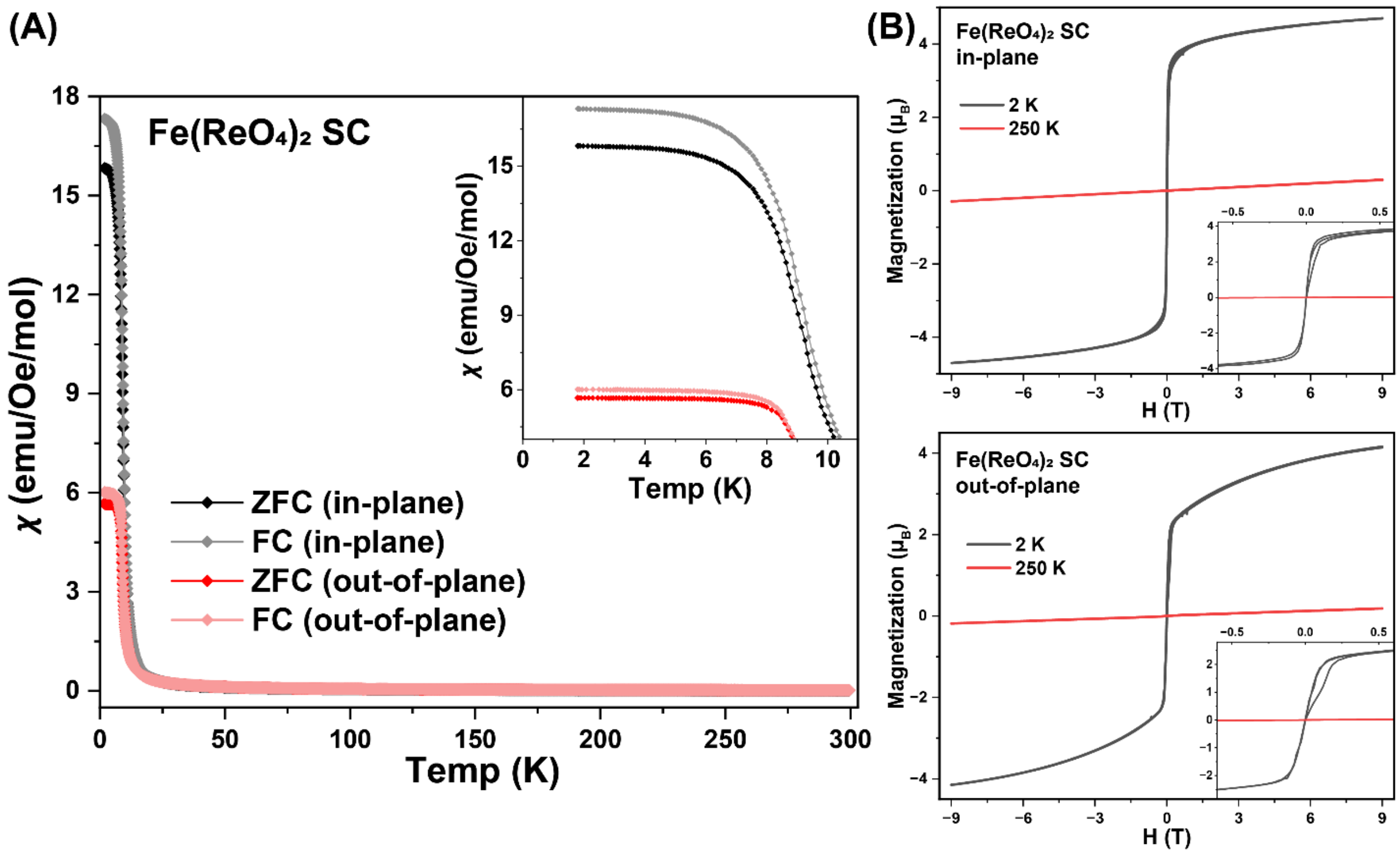

**Figure 9.** (A) The magnetic susceptibility measured from 1.8 to 300 K under 1000 Oe (ZFC & FC) for $Fe(ReO_4)_2$ single crystals; (B) the field-dependent magnetization data collected at 2 K and 250 K, with the field direction in-plane (top) and out-of-plane (bottom).

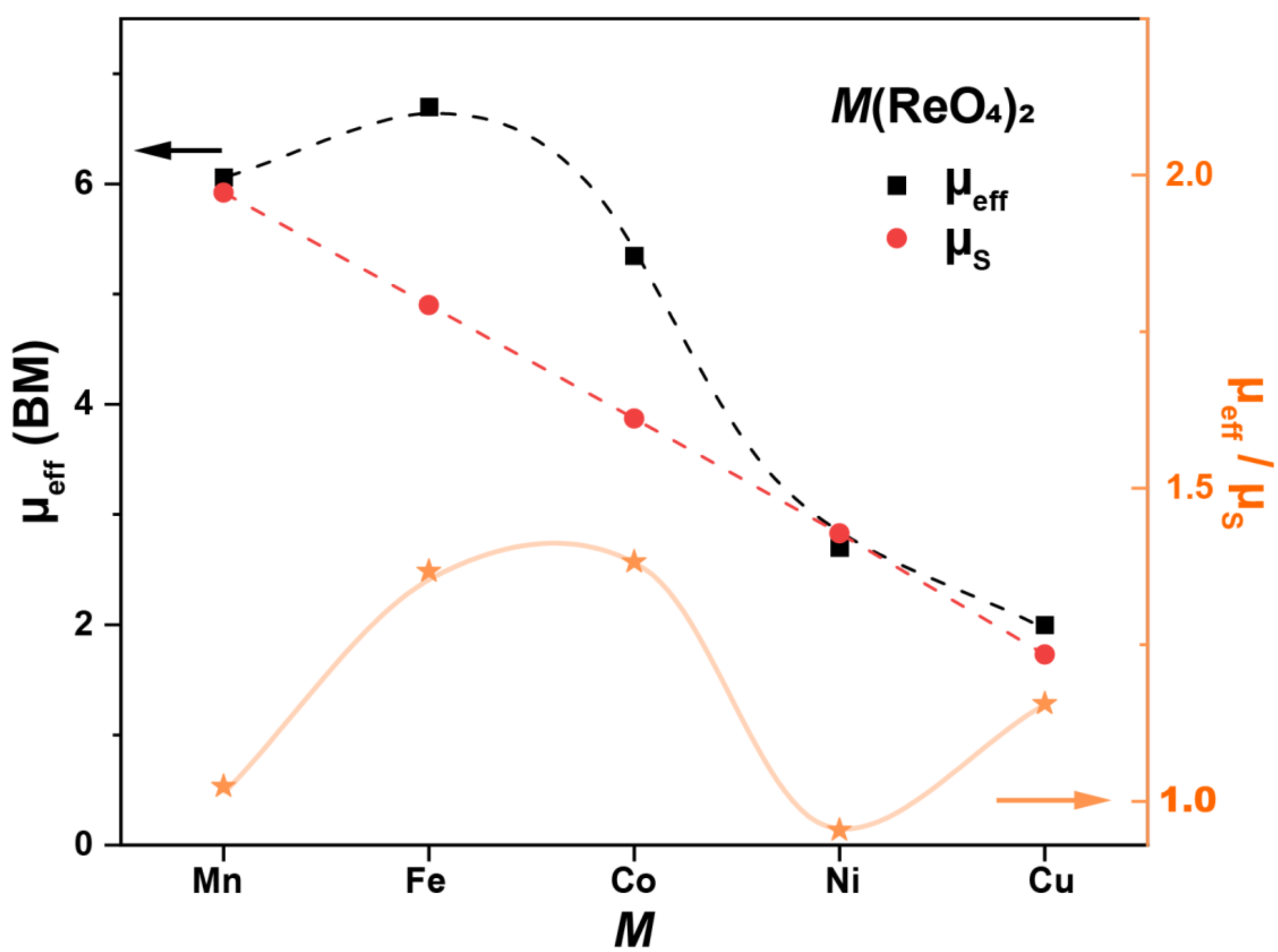

**Figure 10.** Effective moment ($\mu_{eff}$) calculated from the Curie-Weiss fitting of MT data, plotted versus different magnetic transition metal ***M*** of ***M***$(ReO_4)_2$, and compared to the theoretical values of spin-only $\mu_S$. The $\mu_{eff}$ / $\mu_S$ value is plotted on the right y-axis in orange.

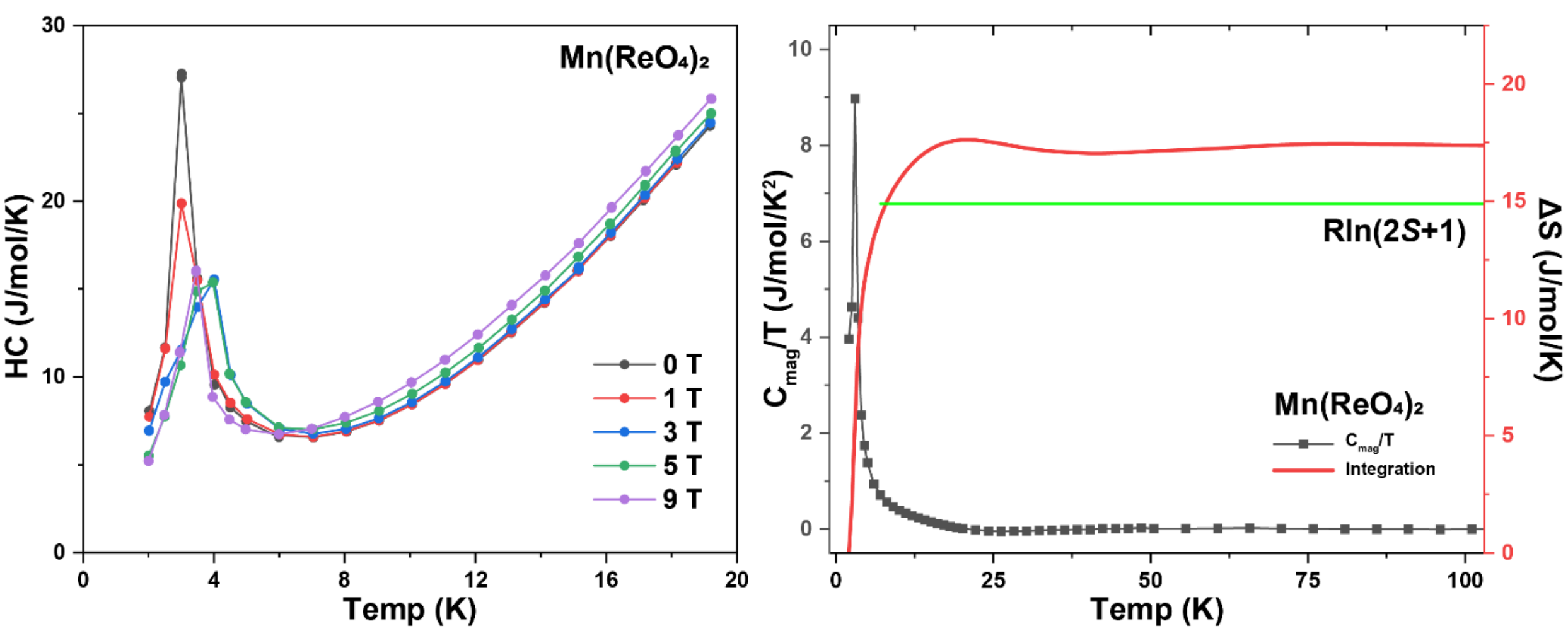

**Figure 11.** The heat capacity (HC) data of a $Mn(ReO_4)_2$ polycrystalline dense piece, measured under different magnetic field (left), together with $C_{mag}/T$ versus T, with ΔS calculated from the integration of $C_{mag}/T$, plotted on the right *y*-axis (right). The green line suggests the theoretical magnetic (spin) entropy of the system using the Heisenberg model (*S* = 5/2).

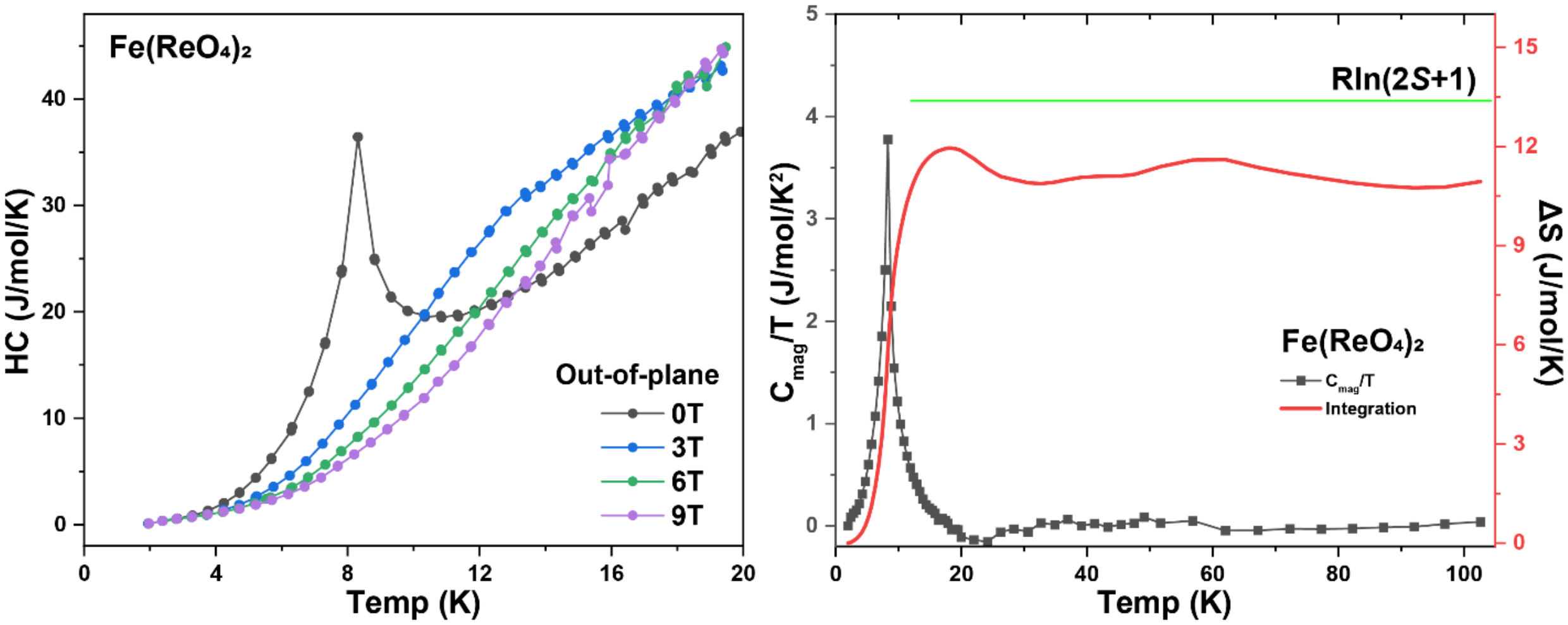

**Figure 12.** The heat capacity (HC) data of a $Fe(ReO_4)_2$ single crystal piece, measured under different magnetic fields (left), together with $C_{mag}/T$ versus T, with ΔS calculated from the integration of $C_{mag}/T$, plotted on the right *y*-axis (right). The green line suggests the theoretical magnetic (spin) entropy of the system using the Heisenberg model (*S* = 2).

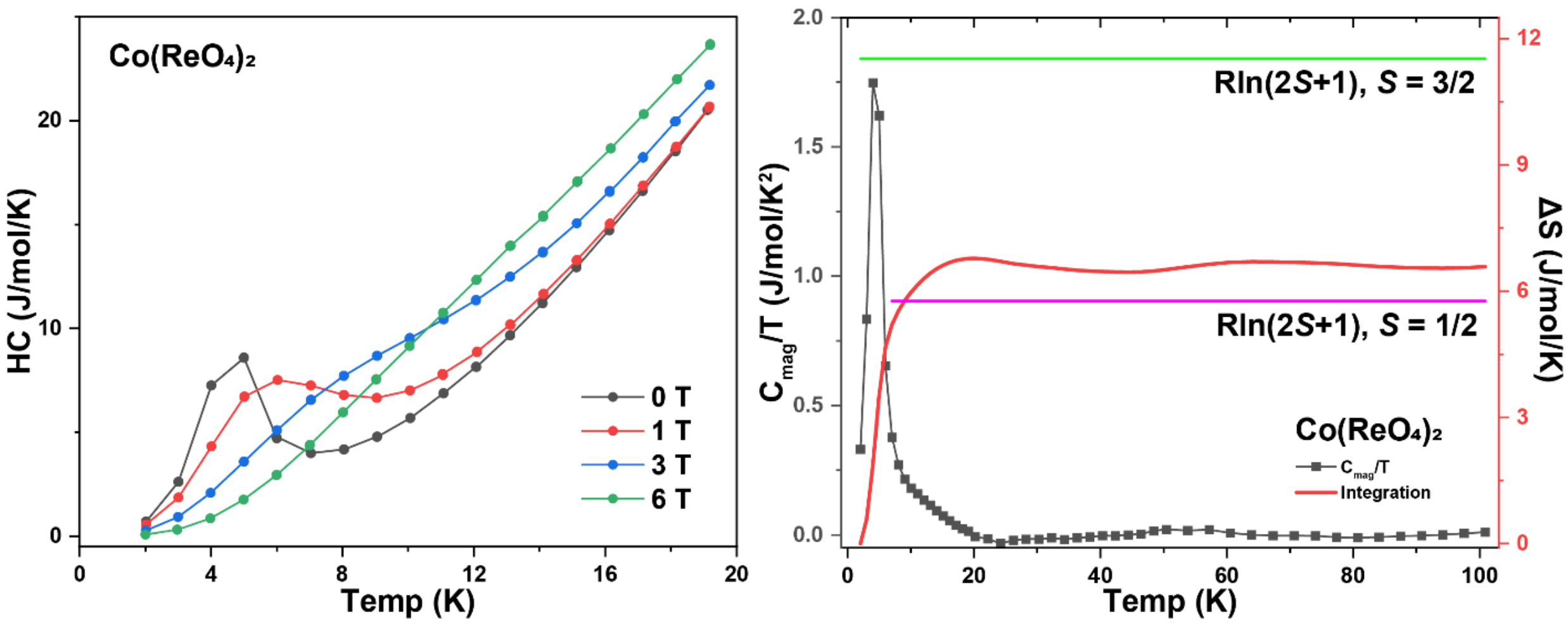

**Figure 13.** The heat capacity (HC) data of a $Co(ReO_4)_2$ polycrystalline dense piece, measured under different magnetic fields (left), together with $C_{mag}/T$ versus T, with ΔS calculated from the integration of $C_{mag}/T$, plotted on the right *y*-axis (right). The green line suggests the theoretical magnetic (spin) entropy of the system using Heisenberg model (*S* = 3/2), while the pink line represents the Ising model (*S* = 1/2).

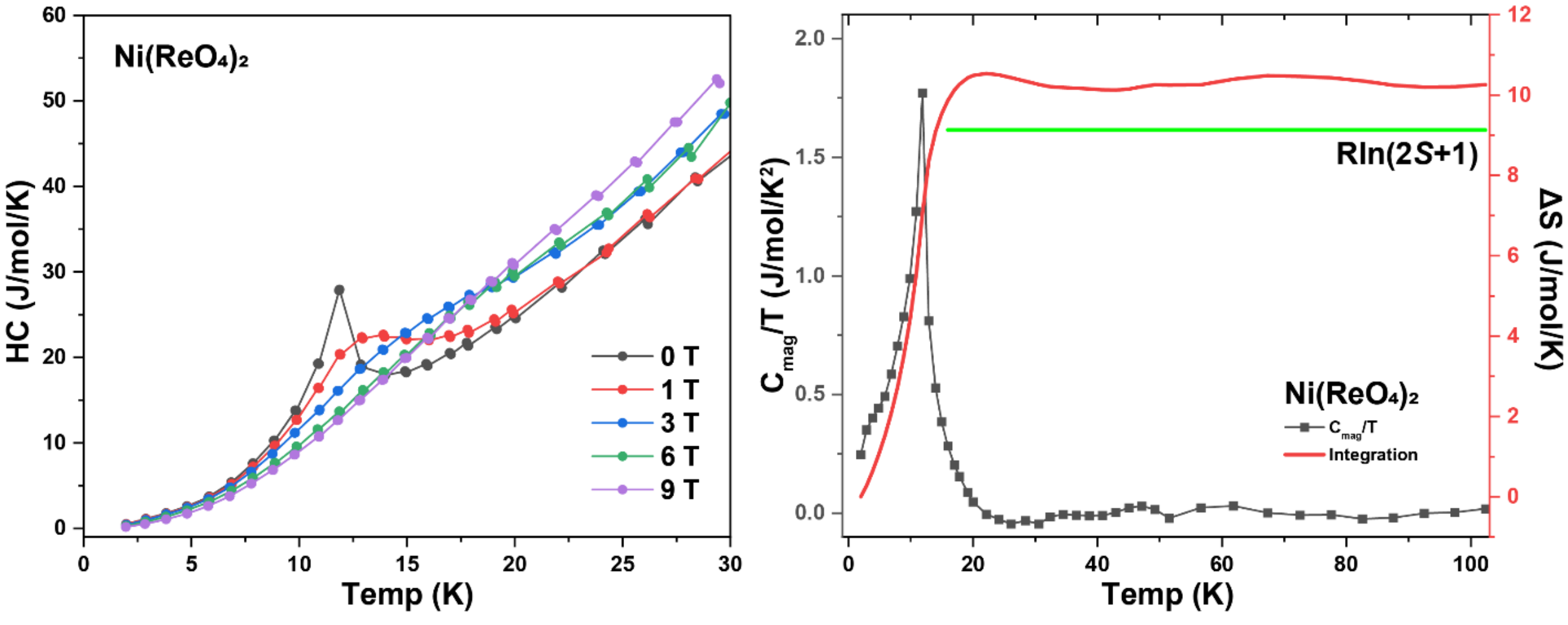

**Figure 14.** The heat capacity (HC) data of a $Ni(ReO_4)_2$ polycrystalline dense piece, measured under different magnetic field (left), together with $C_{mag}/T$ versus T, with ΔS calculated from the integration of $C_{mag}/T$, plotted on the right *y*-axis (right). The green line suggests the theoretical magnetic (spin) entropy of the system using the Heisenberg model (*S* = 1).

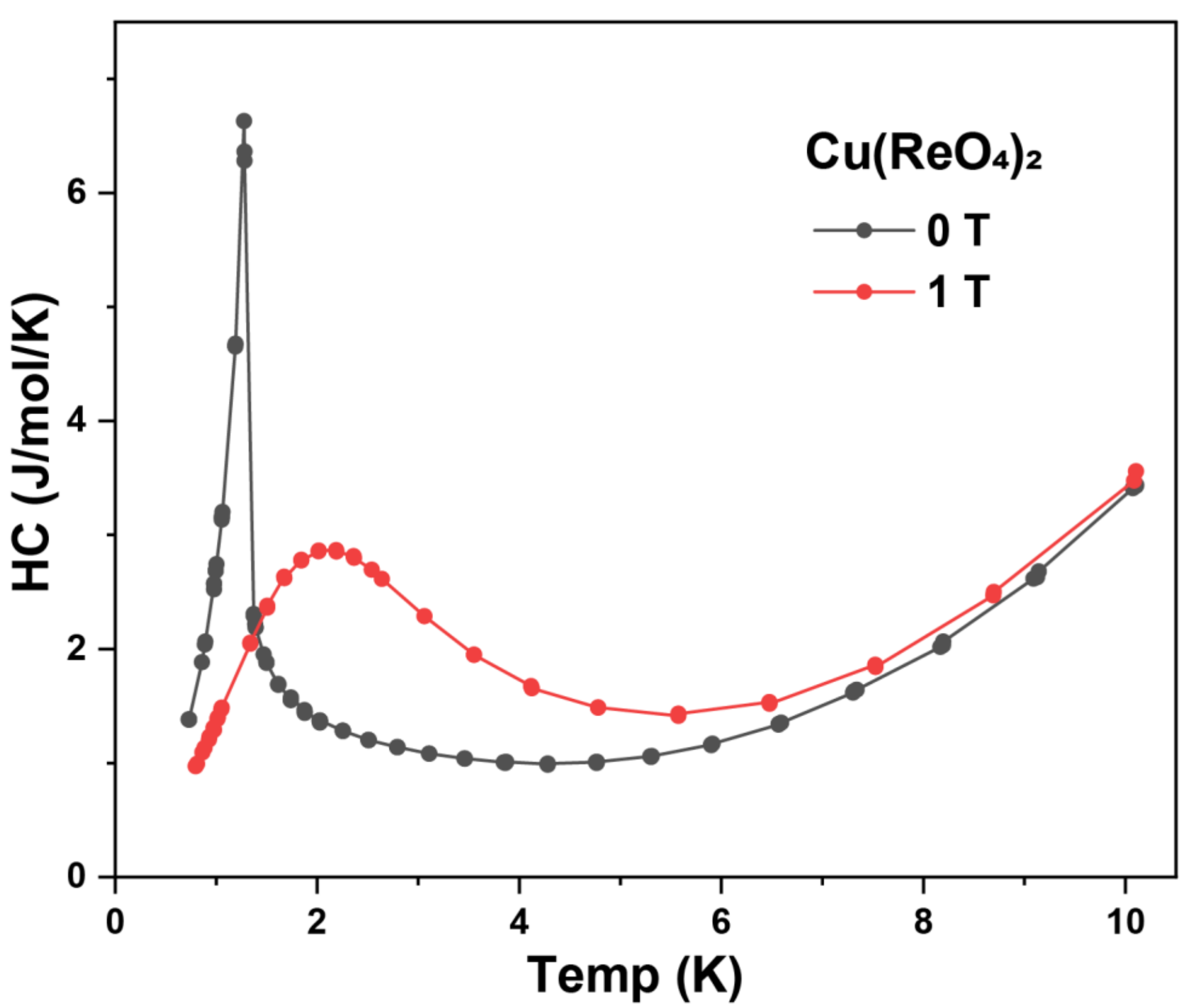

**Figure 15.** The very-low-temperature heat capacity (HC) data for a $Cu(ReO_4)_2$ polycrystalline dense piece, measured under zero magnetic field and 1 T.

*Supporting Information for*

# Crystal structure and basic properties of dirhenate quantum materials

Danrui Ni [a], Xianghan Xu [c], Stephen Zhang [a], N.P. Ong[b], Sanfeng Wu[b] and R. J. Cava* [a]

[a]Department of Chemistry, Princeton University, Princeton, NJ 08544

[b]Department of Physics, Princeton University, Princeton, NJ 08544

[c]School of Physics and Astronomy, University of Minnesota, Twin Cities, MN 55455

*E-mail: rcava@princeton.edu

**This SI includes:**

- Figure S14. Fitting of phonon contribution of the HC data for ***M***$(ReO_4)_2$ (***M*** = Mn, Fe, Co, Ni).

**Supplementary Text**

CCDC deposition number of refined crystal structure

CSD 2558905-2558911 contains the supplemental crystallographic data for this paper. These data can be obtained free of charge (available at www.ccdc.cam.ac.uk/data_request/cif, or by emailing data_request@ccdc.cam.ac.uk, or by contacting The Cambridge Crystallographic Data Centre, 12 Union Road, Cambridge CB2 1EZ, UK, fax: + 441223336033).

**Table S1.** Crystal Data and Structure Refinements for $\boldsymbol{M}(ReO_4)_2$ ($\boldsymbol{M}$ = Mg, Mn, Fe, Co, Ni, Zn) at 295 K. Standard deviations in temperature and cell parameters are indicated by the values in parentheses.

| **Refined Formula** | **$Mg(ReO_4)_2$** | **$Mn(ReO_4)_2$** | **$Fe(ReO_4)_2$** | **$Co(ReO_4)_2$** | **$Ni(ReO_4)_2$** | **$Zn(ReO_4)_2$** |
|---|---|---|---|---|---|---|
| Temperature (K) | 295(2) | 295(2) | 295(2) | 295(2) | 295(2) | 295(2) |
| F.W. (g/mol) | 524.71 | 555.34 | 556.25 | 559.33 | 559.11 | 565.77 |
| Crystal System | Trigonal | Trigonal | Trigonal | Trigonal | Trigonal | Trigonal |
| Space Group | *P*-3*m*1 (164) | *P*-3*m*1 (164) | *P*-3*m*1 (164) | *P*-3*m*1 (164) | *P*-3*m*1 (164) | *P*-3*m*1 (164) |
| Unit Cell Dimension *a* (Å) | 5.7321(1) | 5.8661(1) | 5.7674(1) | 5.7307(2) | 5.6718(2) | 5.7310(1) |
| Unit Cell Dimension *c* (Å) | 6.1611(3) | 6.0755(3) | 6.1259(2) | 6.1197(3) | 6.1407(3) | 6.1592(2) |
| Volume ($Å^3$) | 175.314(12) | 181.055(12) | 176.466(10) | 174.051(15) | 171.076(15) | 175.193(7) |
| *Z* | 1 | 1 | 1 | 1 | 1 | 1 |
| Calculated Density ($g/cm^3$) | 4.970 | 5.093 | 5.234 | 5.336 | 5.427 | 5.363 |
| Absorption Coefficient ($mm^{-1}$) | 34.581 | 35.061 | 36.239 | 37.041 | 38.012 | 37.855 |
| F(000) | 226 | 239 | 240 | 241 | 242 | 244 |
| Theta range for data collection | 3.307° to 26.329° | 3.353° to 27.059° | 3.326° to 26.718° | 3.329° to 26.810° | 3.318° to 25.340° | 3.308° to 26.798° |
| Index Ranges | $-6 \leq h \leq 7$, $-7 \leq k \leq 7$, $-7 \leq l \leq 7$ | $-7 \leq h \leq 7$, $-7 \leq k \leq 7$, $-7 \leq l \leq 7$ | $-7 \leq h \leq 7$, $-7 \leq k \leq 7$, $-7 \leq l \leq 7$ | $-7 \leq h \leq 7$, $-7 \leq k \leq 7$, $-7 \leq l \leq 7$ | $-6 \leq h \leq 6$, $-6 \leq k \leq 6$, $-7 \leq l \leq 7$ | $-7 \leq h \leq 7$, $-7 \leq k \leq 7$, $-7 \leq l \leq 7$ |
| Reflections Collected | 2451 | 2827 | 2867 | 2792 | 1654 | 2957 |
| Independent Reflections | 160 [R(int) = 0.0379] | 177 [R(int) = 0.0379] | 171 [R(int) = 0.0503] | 171 [R(int) = 0.0332] | 140 [R(int) = 0.0371] | 171 [R(int) = 0.0403] |
| Goodness-of-fit on $F^2$ | 1.129 | 1.110 | 1.170 | 1.204 | 1.341 | 1.232 |
| Final R Indices [$I > \sigma(I)$] | R1 = 0.0111, wR2 = 0.0264 | R1 = 0.0115, wR2 = 0.0267 | R1 = 0.0150, wR2 = 0.0286 | R1 = 0.0094, wR2 = 0.0199 | R1 = 0.0154, wR2 = 0.0267 | R1 = 0.0121, wR2 = 0.0275 |
| R Indices (all data) | R1 = 0.0118, wR2 = 0.0265 | R1 = 0.0125, wR2 = 0.0268 | R1 = 0.0174, wR2 = 0.0291 | R1 = 0.0099, wR2 = 0.0199 | R1 = 0.0167, wR2 = 0.0270 | R1 = 0.0128, wR2 = 0.0276 |
| Largest Diff. Peak and Hole (e.A$^{-3}$) | 0.485 and -0.502 | 1.169 and -0.699 | 0.707 and -0.800 | 0.556 and -0.626 | 0.872 and -2.042 | 0.711 and -0.491 |

**Table S2.** Crystal Data and Structure Refinements for ***M***$(ReO_4)_2$ (***M*** = Cu) at 295 K, with selected bond lengths (Å) and bond angles (°). Standard deviations in temperature and cell parameters are indicated by the values in parentheses.

| **Refined Formula** | **$Cu(ReO_4)_2$** | |
|---|---|---|
| Temperature (K) | 295(2) | |
| F.W. (g/mol) | 563.94 | |
| Crystal System | Monoclinic | |
| Space Group | *C*2/*m* (12) | |
| Unit Cell Dimension *a* (Å) | 10.1815(13) | |
| Unit Cell Dimension *b* (Å) | 5.5786(7) | |
| Unit Cell Dimension *c* (Å) | 6.0186(8) | |
| Unit Cell Angle *β* (°) | 92.681(5) | |
| Volume ($Å^3$) | 341.47(8) | |
| *Z* | 2 | |
| Calculated Density ($g/cm^3$) | 5.485 | |
| Absorption Coefficient ($mm^{-1}$) | 38.445 | |
| F(000) | 486 | |
| Theta range for data collection | 3.389° to 26.359° | |
| Index Ranges | $-12 \le h \le 12$,<br>$-6 \le k \le 6$,<br>$-7 \le l \le 7$ | |
| Reflections Collected | 4022 | |
| Independent Reflections | 388 [R(int) = 0.0329] | |
| Goodness-of-fit on $F^2$ | 1.183 | |
| Final R Indices [$I > \sigma(I)$] | R1 = 0.0160, wR2 = 0.0373 | |
| R Indices (all data) | R1 = 0.0166, wR2 = 0.0375 | |
| Largest Diff. Peak and Hole ($e.A^{-3}$) | 0.936 and -1.097 | |
| Selected bond lengths (Å) and bond angles (°) | | |
| **Bond Length (Å)** | Cu-O1 (x4) | 1.943(4) |
| | Cu-O2 (x2) | 2.309(6) |
| | Re-O1 (x2) | 1.731(4) |
| | Re-O2 (x1) | 1.687(6) |
| | Re-O3 (x1) | 1.700(7) |
| **Bond Angle (°)** | O1-Cu-O1 (x2) | 90.1(3) |
| | O1-Cu-O1 (x2) | 89.9(3) |
| | O1-Cu-O2 (x4) | 90.92(18) |
| | O1-Cu-O2 (x4) | 89.08(18) |
| | O1-Re-O2 (x2) | 110.00(19) |
| | O1-Re-O3 (x2) | 108.9(2) |
| | O2-Re-O3 (x1) | 109.5(4) |
| | O1-Re-O1 (x1) | 109.6(3) |
| ***M*O6 octahedron** | **Quadratic elongation** | 1.0140 |
| | **Distortion index (bond length)** | 0.07897 |

**Table S3.** Selected bond lengths (Å) and bond angles (°) for ***M***$(ReO_4)_2$ (***M*** = Mg, Mn, Fe, Co, Ni, Zn, Cu) at 295 K. Standard deviation is indicated by the values in parentheses.

| | **$Mg(ReO_4)_2$** | | **$Mn(ReO_4)_2$** | | **$Fe(ReO_4)_2$** | |
|---|---|---|---|---|---|---|
| **Bond Length (Å)** | Mg-O1 (x6) | 2.058(3) | Mn-O1 (x6) | 2.135(3) | Fe-O1 (x6) | 2.085(4) |
| | Re-O1 (x3) | 1.714(3) | Re-O1 (x3) | 1.715(3) | Re-O1 (x3) | 1.709(4) |
| | Re-O2 (x1) | 1.699(6) | Re-O2 (x1) | 1.691(7) | Re-O2 (x1) | 1.694(7) |
| **Bond Angle (°)** | O1-Mg-O1 (x6) | 89.40(14) | O1-Mn-O1 (x6) | 88.58(13) | O1-Fe-O1 (x6) | 88.90(17) |
| | O1-Mg-O1 (x6) | 90.60(14) | O1-Mn-O1 (x6) | 91.42(13) | O1-Fe-O1 (x6) | 91.10(17) |
| | O1-Re-O2 (x3) | 109.08(12) | O1-Re-O2 (x3) | 108.95(12) | O1-Re-O2 (x3) | 109.42(15) |
| | O1-Re-O1 (x3) | 109.86(12) | O1-Re-O1 (x3) | 109.99(12) | O1-Re-O1 (x3) | 109.52(15) |
| ***M*O6 octahedron** | Quadratic elongation | 1.0001 | Quadratic elongation | 1.0006 | Quadratic elongation | 1.0004 |
| | **$Co(ReO_4)_2$** | | **$Ni(ReO_4)_2$** | | **$Zn(ReO_4)_2$** | |

| **Bond Length (Å)** | Co-O1 (x6) | 2.056(3) | Ni-O1 (x6) | 2.018(5) | Zn-O1 (x6) | 2.060(4) |
|---|---|---|---|---|---|---|
| | Re-O1 (x3) | 1.713(3) | Re-O1 (x3) | 1.717(5) | Re-O1 (x3) | 1.712(4) |
| | Re-O2 (x1) | 1.703(5) | Re-O2 (x1) | 1.701(9) | Re-O2 (x1) | 1.691(8) |
| **Bond Angle (°)** | O1-Co-O1 (x6) | 89.16(13) | O1-Ni-O1 (x6) | 89.5(2) | O1-Mg-O1 (x6) | 89.28(19) |
| | O1-Co-O1 (x6) | 90.84(13) | O1-Ni-O1 (x6) | 90.5(2) | O1-Mg-O1 (x6) | 90.72(19) |
| | O1-Re-O2 (x3) | 109.18(11) | O1-Re-O2 (x3) | 109.3(2) | O1-Re-O2 (x3) | 109.29(17) |
| | O1-Re-O1 (x3) | 109.76(11) | O1-Re-O1 (x3) | 109.6(2) | O1-Re-O1 (x3) | 109.65(17) |
| ***M*O6 octahedron** | **Quadratic elongation** | 1.0002 | **Quadratic elongation** | 1.0001 | **Quadratic elongation** | 1.0002 |

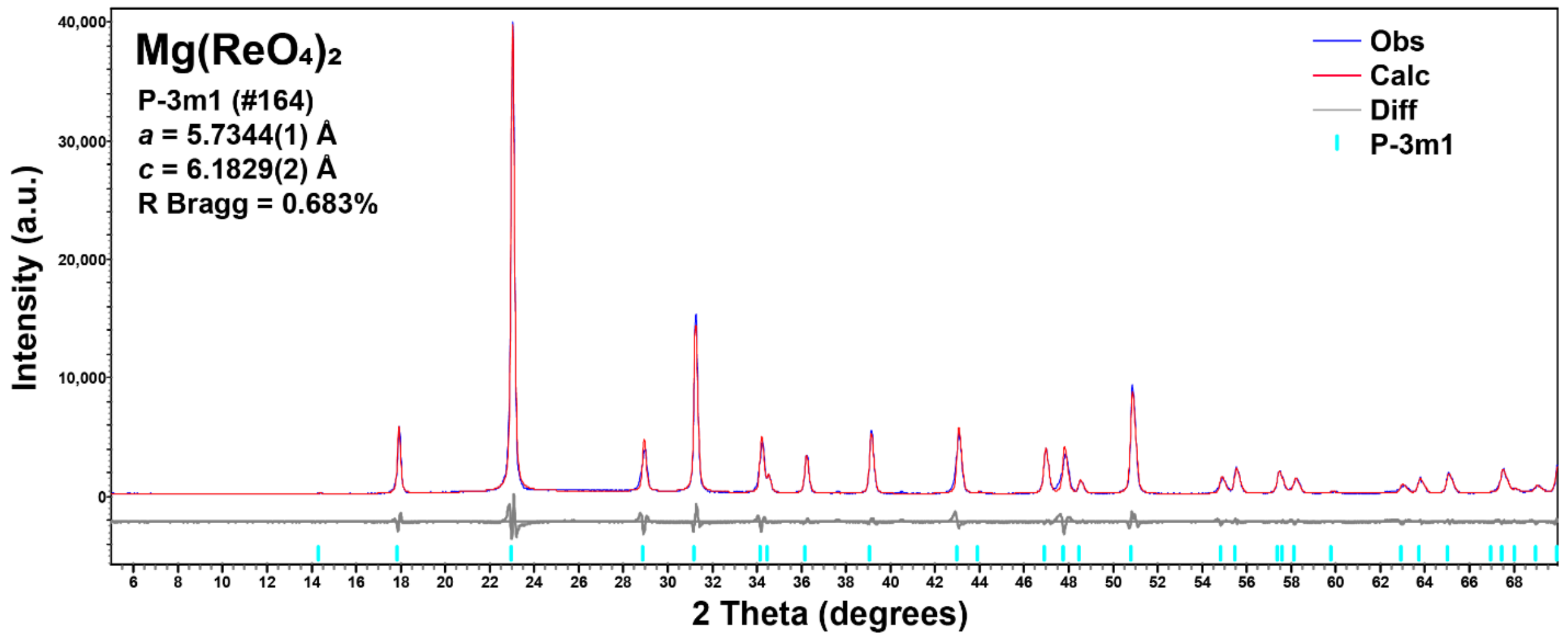

**Figure S1.** PXRD patterns with Le Bail fit for bulk $Mg(ReO_4)_2$, confirming the consistency of the bulk sample structure and composition with the SCXRD refinement results. R Bragg = 0.683%.

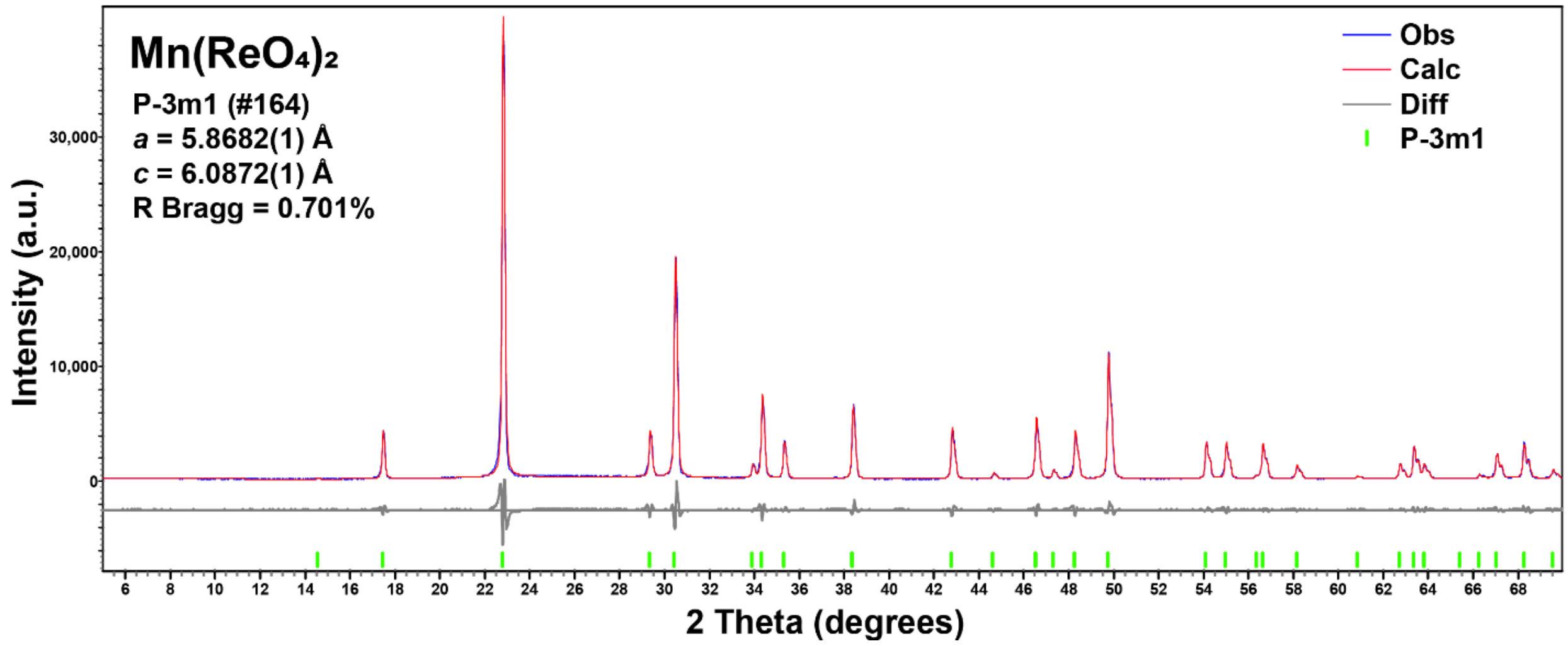

**Figure S2.** PXRD patterns with Le Bail fit for bulk $Mn(ReO_4)_2$, confirming the consistency of the bulk sample structure and composition with the SCXRD refinement results. R Bragg = 0.701%.

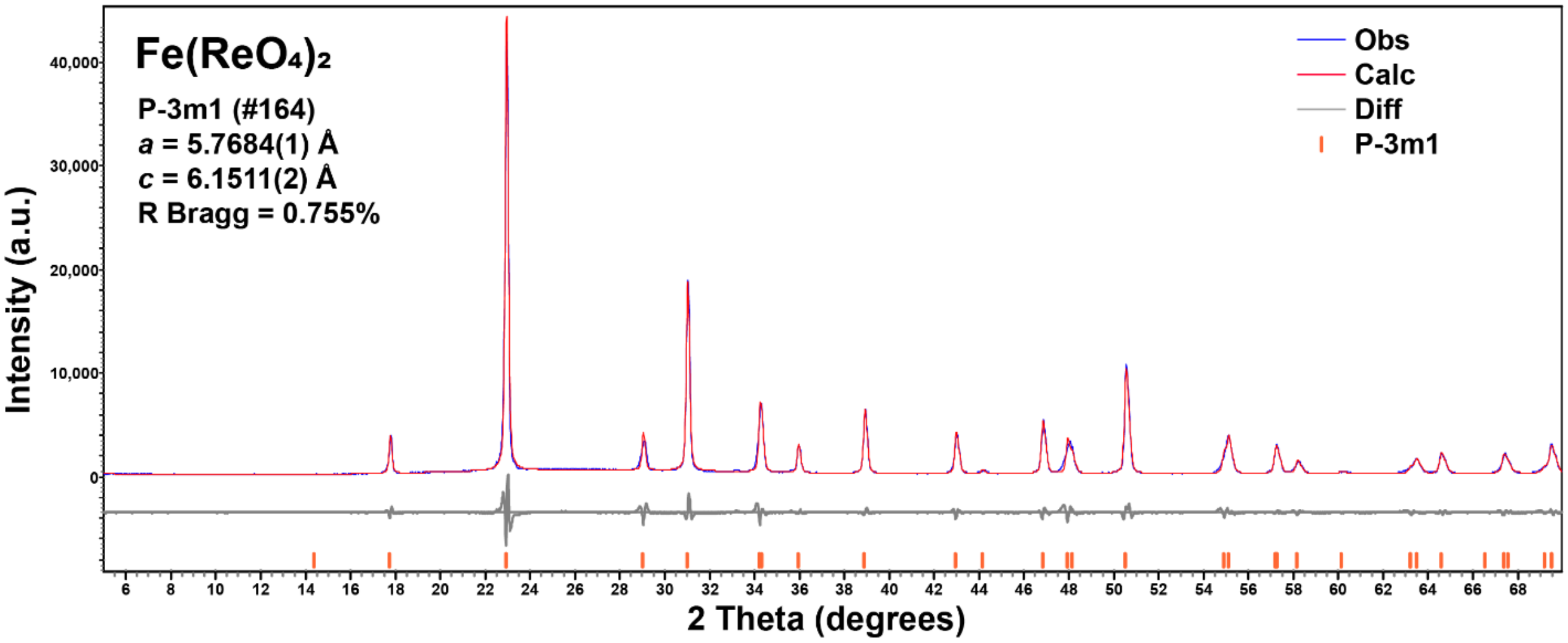

**Figure S3.** PXRD patterns with Le Bail fit for bulk $Fe(ReO_4)_2$, confirming the consistency of the bulk sample structure and composition with the SCXRD refinement results. R Bragg = 0.755%.

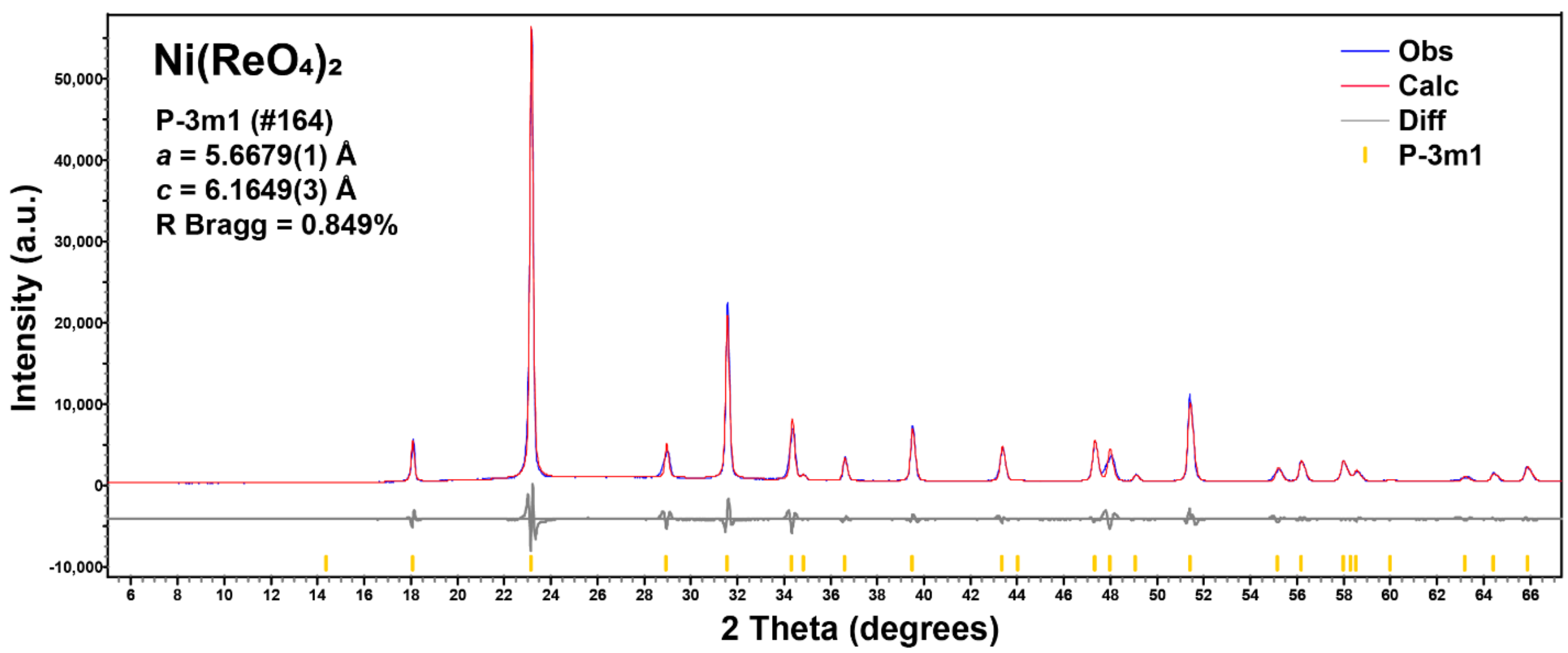

**Figure S4.** PXRD patterns with Le Bail fit for bulk $Ni(ReO_4)_2$, confirming the consistency of the bulk sample structure and composition with the SCXRD refinement results. R Bragg = 0.849%.

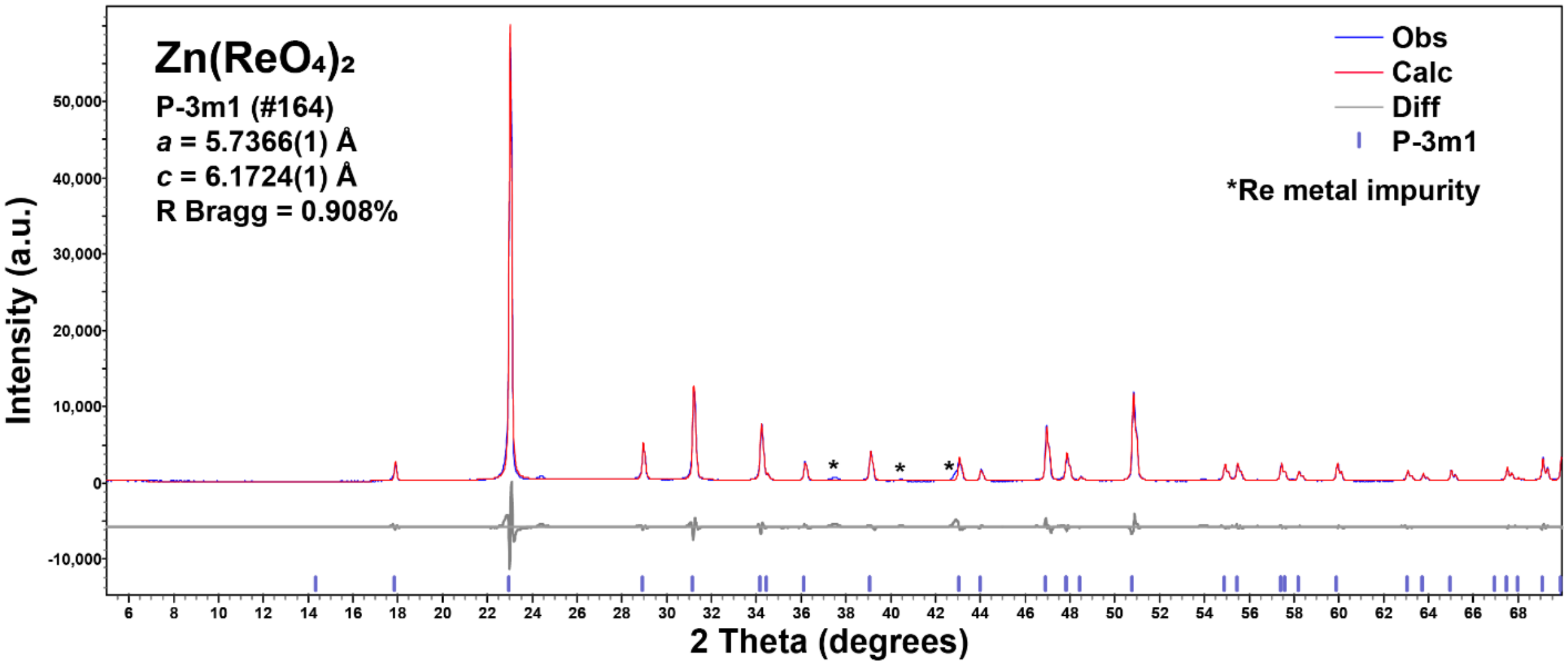

**Figure S5.** PXRD patterns with Le Bail fit for bulk $Zn(ReO_4)_2$, confirming the consistency of the bulk sample structure and composition with the SCXRD refinement results. R Bragg = 0.908%.

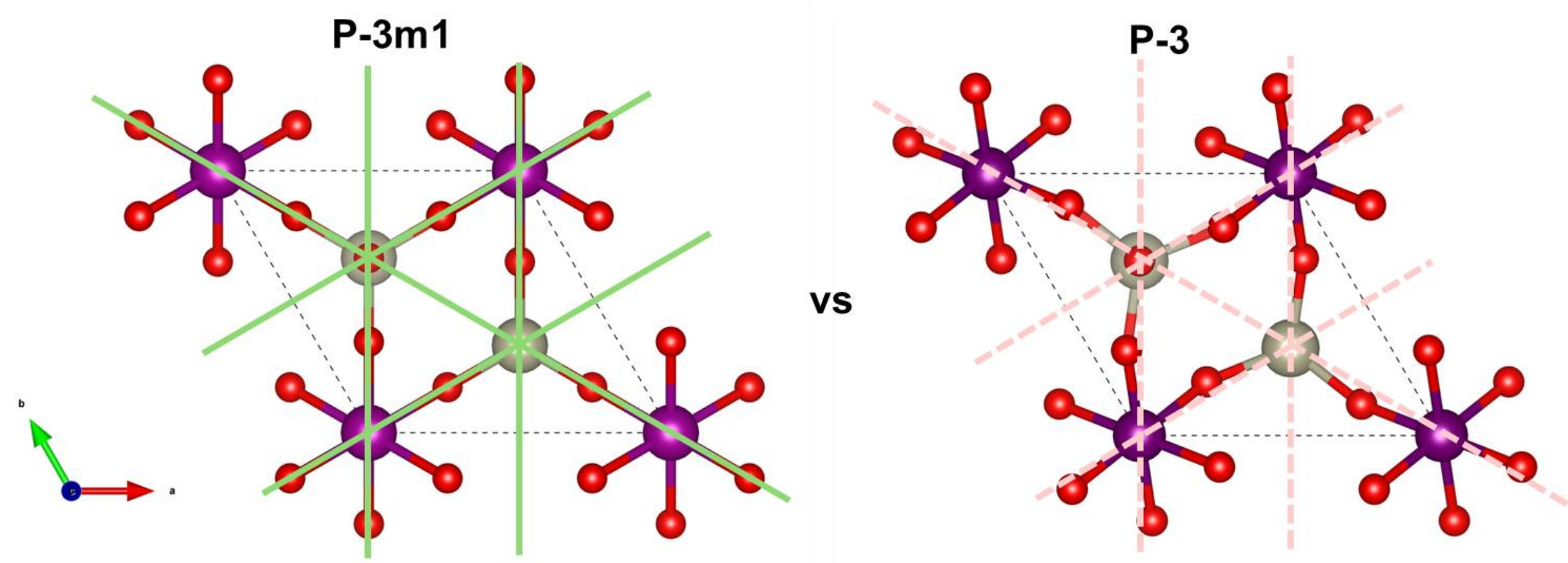

**Figure S6.** Comparison of the structure of *P*-3*m*1 (left) and *P*-3 (right[1]) unit cell in the view along *c*-axis. The colored lines help to find the mirror positions of *P*-3*m*1, and label out how the oxygen in *P*-3 breaks the mirror symmetry.

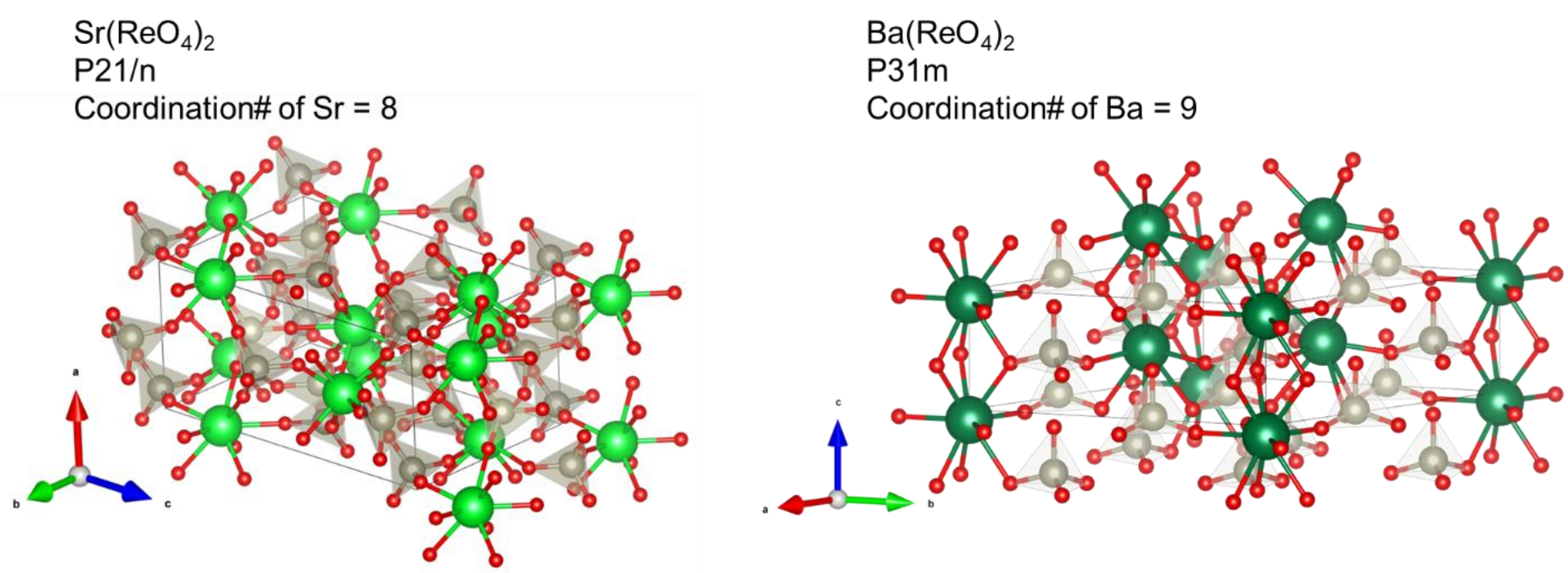

**Figure S7.** The unit cell of $Sr(ReO_4)_2$ (left[2]) and $Ba(ReO_4)_2$ (right[3,4]). With larger divalent metal ions, its coordination number increases, crystallizing into a different structure of rhenates.

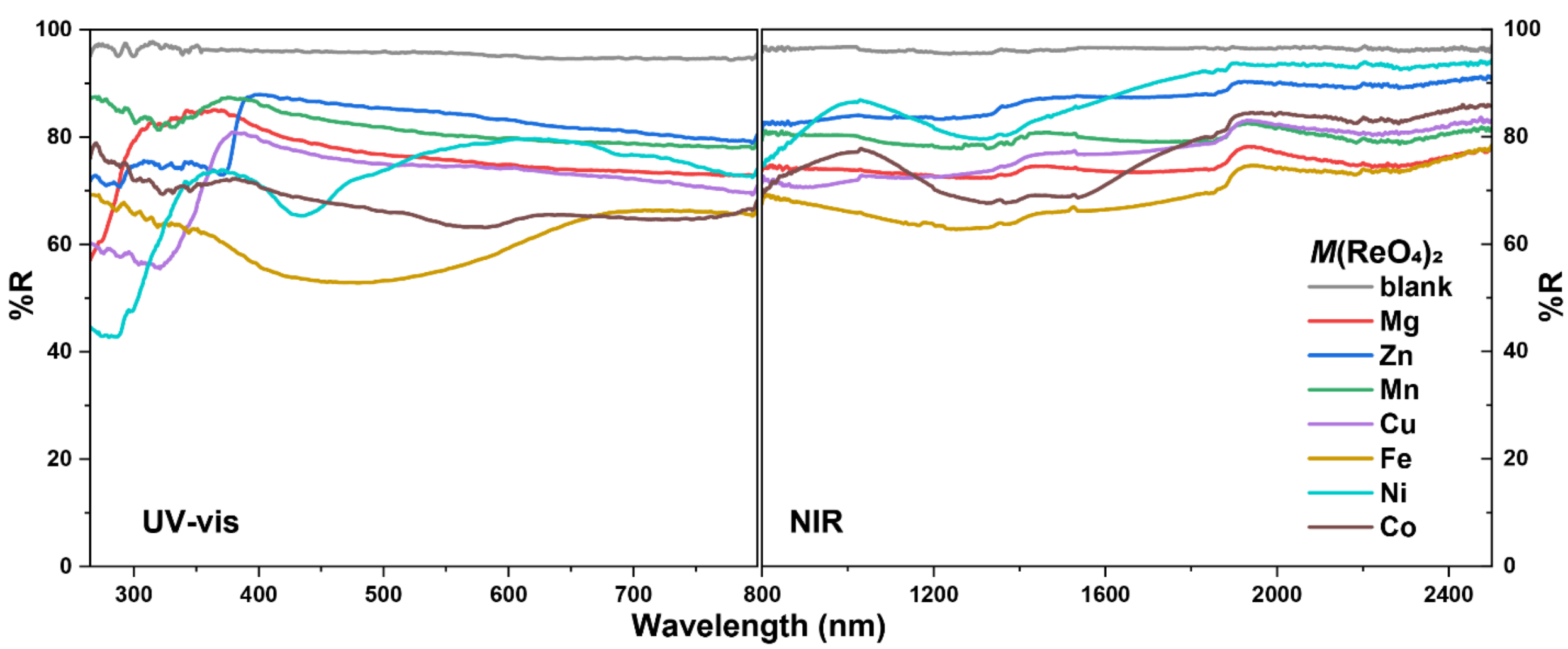

**Figure S8.** Diffuse reflectance (DR) spectra of polycrystalline $\boldsymbol{M}(ReO_4)_2$ powder samples, collected from 250 to 2500 nm.

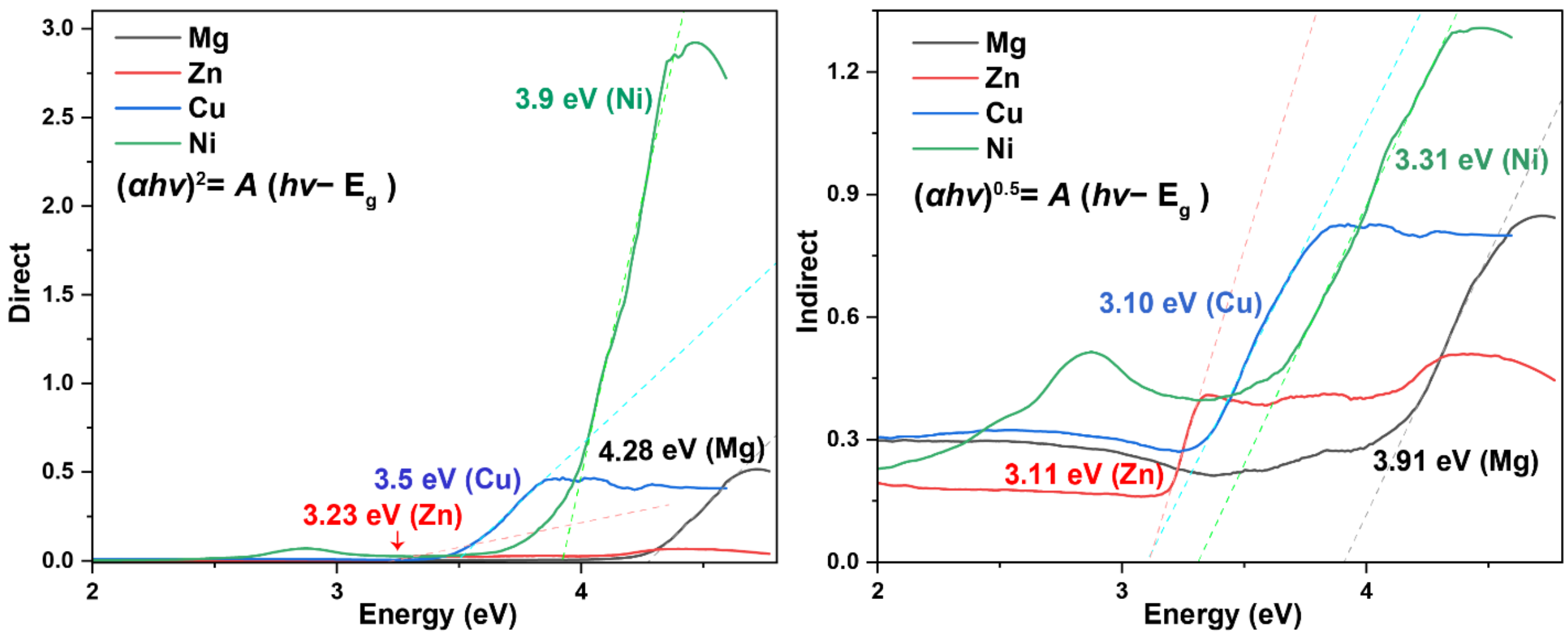

**Figure S9.** The Tauc plots of selected $\mathbf{M}(ReO_4)_2$ samples ($\boldsymbol{M}$ = Ni/Cu/Mg/Zn), with direct (left) and indirect (right) transition equations.

The reflectance data were converted to absorption by using the Kubelka–Munk function and the band gap values are estimated using Tauc plots[5,6]:

$$(\alpha h\nu)^n = A\left(h\nu - E_g\right)$$

where $n$ = 2 for direct transition and 0.5 for indirect. $A$ is a constant and $\alpha$ is the absorption coefficient ($cm^{-1}$).

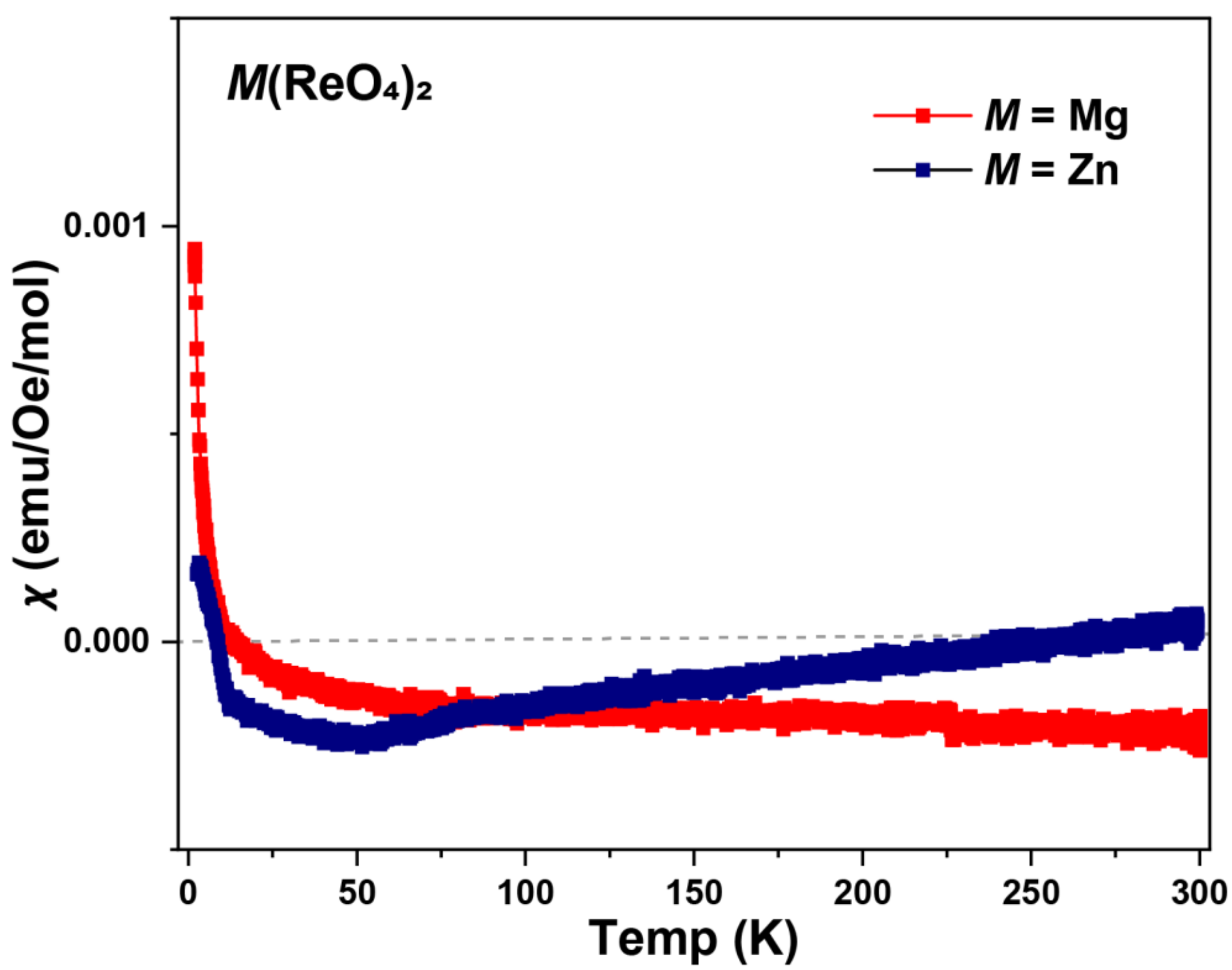

**Figure S10.** The magnetic susceptibility measured from 1.8 to 300 K under 1000 Oe (ZFC) for ***M***$(ReO_4)_2$ polycrystalline powders, with ***M*** = Mg and Zn.

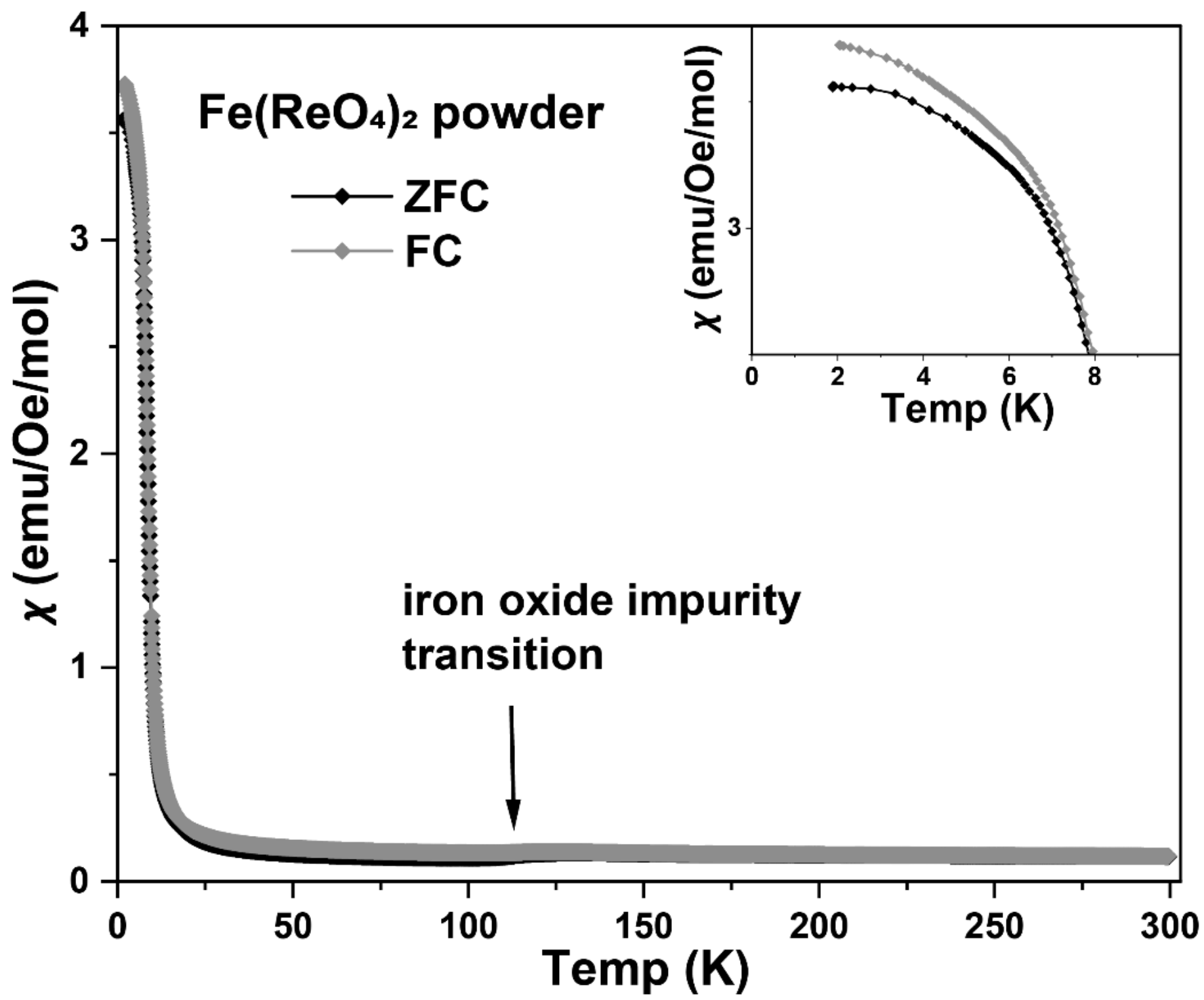

**Figure S11.** The magnetic susceptibility measured from 1.8 to 300 K under 1000 Oe (ZFC & FC) for $\boldsymbol{M}(ReO_4)_2$ polycrystalline powders, with $\boldsymbol{M}$ = Fe.

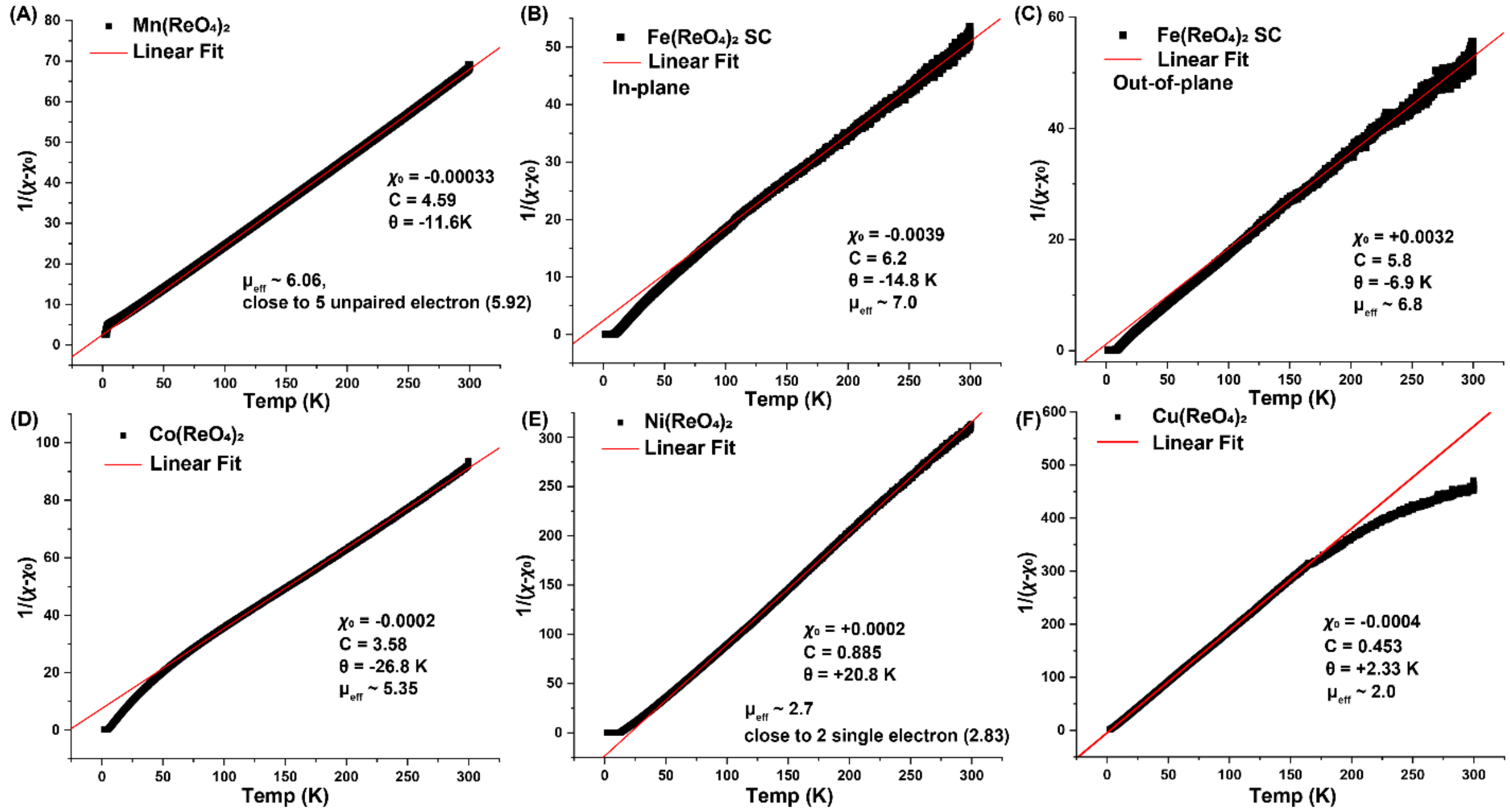

**Figure S12**. Curie-Weiss fitting ($\chi - \chi_0 = \frac{C}{T-\theta}$) conducted in the uninterfered range of the ZFC magnetic susceptibility data measured under 1000 Oe for $\boldsymbol{M}(ReO_4)_2$ samples, with $\boldsymbol{M}$ = (A) Mn, (B) Fe (single crystal in-plane), (C) Fe (single crystal out-of-plane), (D) Co, (E) Ni, and (F) Cu.

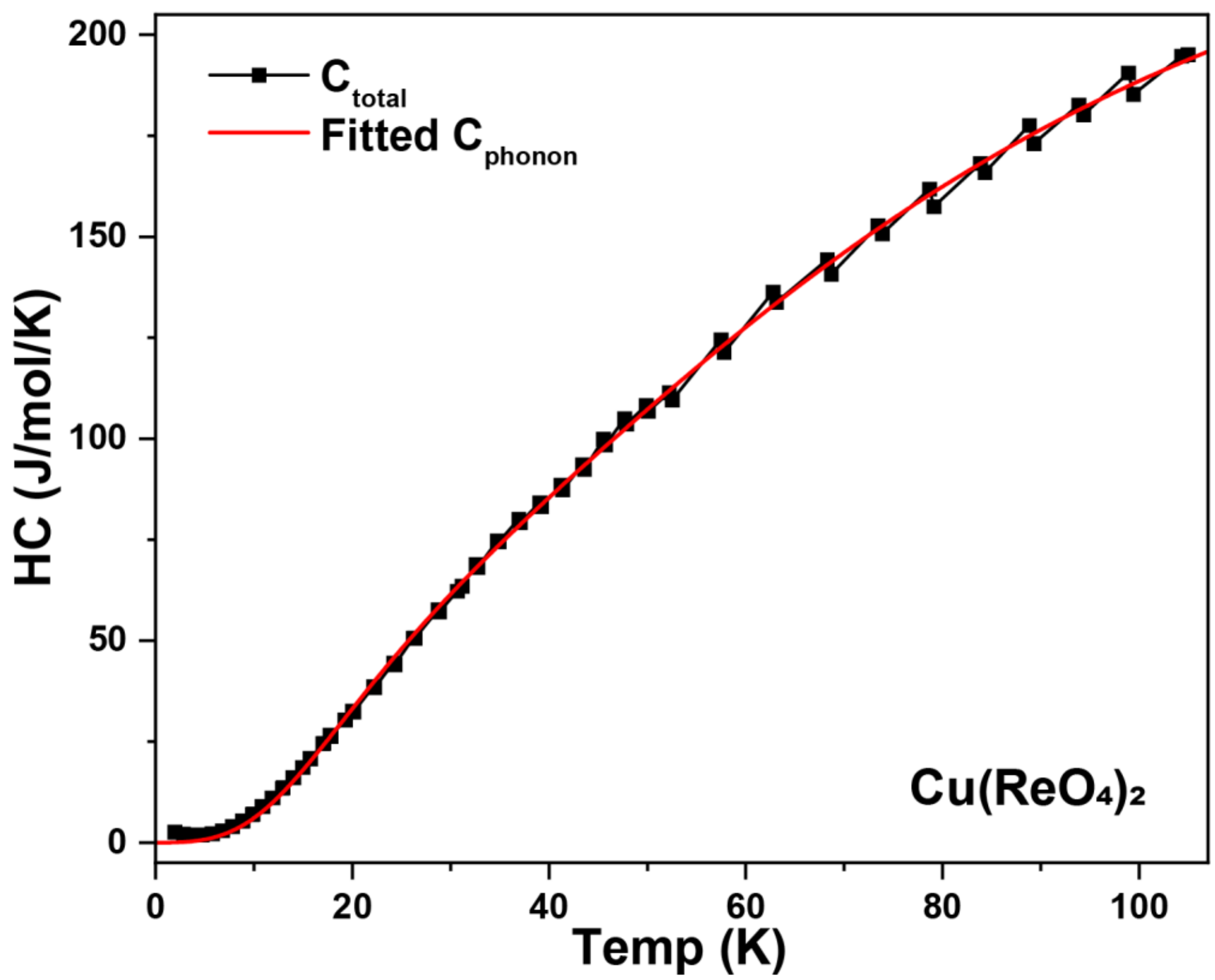

**Figure S13.** Heat capacity data of an as-made polycrystalline piece of $Cu(ReO_4)_2$, measured under zero magnetic field from 2 to 100 K.

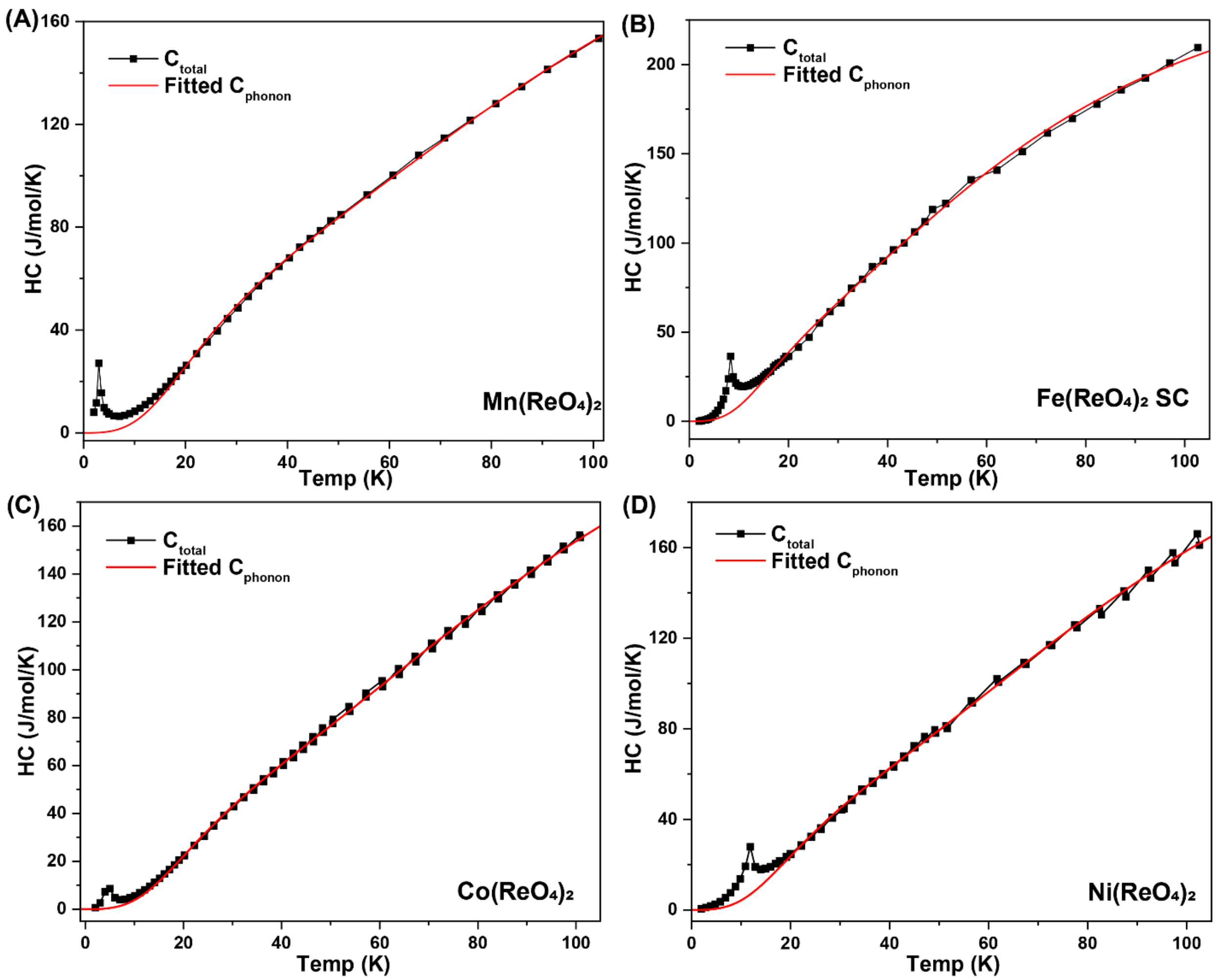

**Figure S14.** Fitting of the phonon contribution in the total heat capacity for ***M***$(ReO_4)_2$ samples, with ***M*** = (A) Mn, (B) Fe (single crystal out-of-plane), (C) Co, and (D) Ni, using modified Debye equation[7] $C_{phonon} = 9R\sum_{n=1}^{2} C_n \left(\frac{T}{\Theta_{Dn}}\right)^3 \int_0^{\Theta_{Dn}/T} \frac{x^4 e^x}{(e^x-1)^2} dx$, in the temperature range without potential magnetic contributions.